\documentclass[twocolumn]{aa}  

\usepackage{graphicx}
\usepackage{txfonts}
\usepackage{xspace}
\usepackage{placeins}
\usepackage{natbib}
\usepackage{xcolor}
\usepackage{amsmath}
\usepackage{amssymb}
\usepackage[colorlinks=true,linkcolor=blue,citecolor=blue,urlcolor=blue]{hyperref}

\newcommand{\lsun}{\mbox{L}_\odot}
\newcommand{\msun}{\mbox{M}_\odot}
\newcommand{\macc}{\dot{M}_{\rm acc}}

\newcommand\kms{km\,s$^{-1}$}
\newcommand{\ArII}{[\ion{Ar}{ii}]}
\newcommand{\OII}{[\ion{O}{ii}]}
\newcommand{\CaII}{[\ion{Ca}{ii}]}
\newcommand{\ClI}{[\ion{Cl}{i}]}
\newcommand{\FeI}{[\ion{Fe}{i}]}
\newcommand{\FeII}{[\ion{Fe}{ii}]} 
\newcommand{\NII}{[\ion{N}{ii}]}
\newcommand{\NiII}{[\ion{Ni}{ii}]}
\newcommand{\NeII}{[\ion{Ne}{ii}]}
\newcommand{\NeIII}{[\ion{Ne}{iii}]}
\newcommand{\SI}{[\ion{S}{i}]}
\newcommand{\SII}{[\ion{S}{ii}]}

\begin{document} 

   \title{JOYS: JWST MIRI/MRS spectra of the inner 500\,au region of the L1527 IRS bipolar outflow}
    \titlerunning{JWST MIRI spectra of L1527 bipolar outflow}
    \authorrunning{Devaraj et al. }

\author{R. Devaraj, \inst{1}, E. F. van Dishoeck \inst{2}, T. P. Ray \inst{1}, \L{}. Tychoniec \inst{2}, A. Caratti o Garatti \inst{3}, L. Francis \inst{2}, C. Gieser \inst{4}, M. L. van Gelder \inst{2}, J. J. Tobin \inst{5}, H. Beuther \inst{4}, P. J. Kavanagh \inst{6}, K. Justtanont \inst{7}, W. B. Drechsler \inst{8}, M. G. Navarro \inst{9} \and G. Perotti \inst{4} }
 
\institute{ $^1$ School of Cosmic Physics, Dublin Institute for Advanced Studies, 31 Fitzwilliam Place, Dublin 2, Ireland\\ 
    $^2$ Leiden Observatory, Leiden University, P.O. Box 9513, NL 2300 RA Leiden, The Netherlands\\
    $^3$ INAF-Osservatorio Astronomico di Capodimonte, Salita Moiariello 16, I-80131 Napoli, Italy\\
    $^4$ Max Planck Institute for Astronomy, K\"{o}nigstuhl 17, 69117 Heidelberg, Germany\\
    $^5$ National Radio Astronomy Observatory, 520 Edgemont Road, Charlottesville, VA 22903, USA\\ 
    $^6$ Department of Physics, Maynooth University, Maynooth, Co. Kildare, Ireland\\
    $^7$ Department of Space, Earth and Environment, Chalmers University of Technology, Onsala Space Observatory, 43992 Onsala, Sweden\\ 
    $^8$ Department of Astronomy, University of Virginia, Charlottesville, VA 22904, USA\\ 
     $^9$ INAF-Osservatorio Astronomico di Roma, Via di Frascati 33, 00078 Monte Porzio Catone, Italy
   }

   \date{Version of \today}

  \abstract
  %\ {} leave it empty if necessary  
   {Outflows and jets are defining characteristics in protostellar evolution, intimately linked to accretion. Understanding their properties and origins is essential for probing the earliest phases of star formation.}
   {This study characterizes the physical and kinematic properties within the innermost 500\,au region of the L1527 bipolar outflow, a class 0/I low-mass protostar, as part of the JWST Observations of Young protoStars (JOYS) program.}
   {We obtained spectroscopic observations using the JWST MIRI/MRS instrument across $5-28\,\mu$m at $0.2-1.0\arcsec$ spatial resolution. We identified emission lines from molecular and ionized species and analyzed their spatial morphology using line-integrated intensity maps. We derived gas temperatures and column densities from excitation diagram analysis of H$_{2}$ rotational lines and compared results with shock models.}
   {The observations reveal extended molecular hydrogen emission tracing the bipolar outflow, with the H$_{2}$ gas temperatures distributed into warm ($\sim$550\,K) and hot ($\sim$2500\,K) components, likely originating from moderate-velocity $J$-type shocks and some UV irradiation. We detect forbidden atomic and ionized emission lines of \NiII, \ArII, \NeII, \NeIII, \SI, and \FeII\ showing spatially extended morphology. Double-peaked emission profiles were seen in \ArII, \NeIII, and \FeII\ in the eastern region, suggesting that the high-velocity component traces a fast, highly ionized jet. A radial velocity map derived from \NeII\ emission shows the eastern region to be redshifted and the western region blueshifted, contrary to earlier interpretations.} 
  {The analysis of the MIRI/MRS observations reveals molecular, atomic, and ionized emission lines in this low-mass protostar connected with active outflow signatures. The most striking feature discovered is a poorly collimated high-velocity ionized jet, embedded within a broader, wide-angle molecular outflow likely driven by a disk wind. The coexistence of these components supports a stratified outflow structure and suggests that L1527 exhibits jet-launching characteristics atypical of its early evolutionary stage.}
   

\maketitle

\begin{figure*}[ht]
\begin{center}
\includegraphics[width=0.85\textwidth]{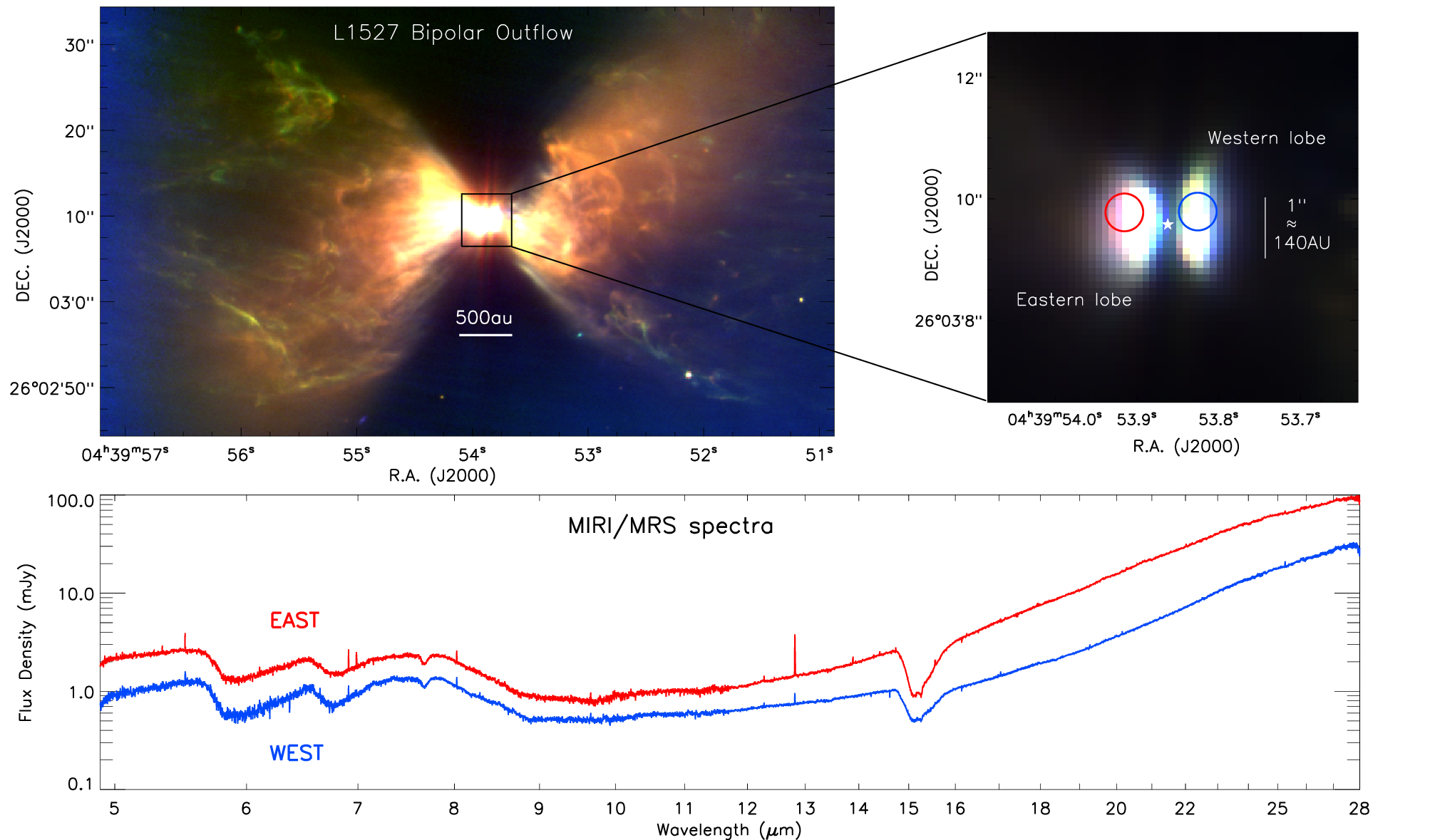}
 \caption{Mid-infrared three-color image of the L1527 bipolar outflow, using JWST MIRI broadband imaging at $5.6\,\mu$m (red), $7.7\,\mu$m (green), and $12.8\,\mu$m (blue). The region spans $1.43\arcmin \times 0.83\arcmin$, corresponding to a physical scale of $12000\,\mathrm{au}\times 7000\,\mathrm{au}$ at a distance of $140\,\mathrm{pc}$. The prominent blue emission at the left edge is not physical, but an artifact of color stretch. Right: Zoomed-in view of the 500\,au central region with brightness scaled to reveal the bipolar outflow lobes. The position of the forming protostar is marked by a star symbol. Two circular apertures of $\sim0\farcs3$ radius in red and blue correspond to the area where the MIRI/MRS spectra were extracted. Bottom: Spectra obtained at the aperture positions across the full MIRI/MRS wavelength range ($5-28\,\mu$m). The red-colored spectra correspond to the eastern lobe, whereas the blue spectra correspond to the western lobe. Flux density of both spectra are scaled for visual offset.}
\label{fig1}
\end{center}
\end{figure*}

\section{Introduction} \label{intro}

The earliest phases of protostar evolution are characterized by accretion and ejection processes, with the forming protostar deeply embedded in its natal envelope. These ejections typically show two connected but distinct observable components: 1) a highly collimated, high-velocity (HV) jet (up to several hundred kilometers per second) whose composition varies with evolutionary stage and may include molecular, atomic, and partially ionized gas, and 2) a slower, wider-angle outflow ($1-50\,$\kms), composed predominantly of molecular and/or atomic gas produced as the jet or a wide-angle wind entrains, shocks, and accelerates the surrounding material \citep{frank14, bally16, ray21}. The exact mechanism driving these outflows is debated, but the main process involves the interaction of accreted matter with the stellar and/or disk magnetic field. Two main models describe the ejection process, depending on the distance from the protostar at which the gas is launched. In the X-wind models \citep{shu07,shang07}, the launch occurs very close to the protostar where the stellar magnetosphere intersects with the protostellar disk. In the disk-wind models, the winds are launched from a few au from the protostar along the magnetized disk surface \citep{blan82,shu94}. In both cases, the poloidal magnetic field plays a critical role in the initial launch and collimation of the outflow \citep{ban06,ban07}. Outflows are now understood to be the primary mechanism in the removal of angular momentum, thereby assisting in accretion \citep{ray07, bally16}. Recent studies suggest that the slower and wider outflow wind influences the evolution and dispersal of protoplanetary disks \citep{pascucci23}. Outﬂows also play a vital role in limiting star formation efficiency and shaping initial mass function \citep{federrath15, guszejnov22}. They can inject energy and momentum into the surrounding medium, dispersing a significant amount of the envelope \citep{dunham14, offner14, zhang16} and thereby ending the infall phase of the protostar. 

Outflows provide insights into their associated protostars and can reveal the nature of their origins through their morphologies, kinematics, and physical properties. Outflows can be probed using a wide array of tracers from visual to radio wavelengths. The atomic and ionized components of outflows have been traced in visual and near-infrared (NIR) hydrogen recombination lines \citep{davis11,nisini24}, and in many forbidden emission transitions such as \OII, \SII, \NII, \NiII, \CaII, and \FeII\ \citep{nisini05,pascucci20}. The molecular component of outflows have been extensively studied in infrared and (sub)millimeter wavelengths using the low-J pure rotational transitions of H$_{2}$ and CO \citep{giannini02,lee02,wu04,davis10,mottram17, dev23}. Accelerated warm and hot gases from shocks have been traced in higher-J transitions of CO \citep{giannini01,gomez12,kristensen17,karska18} and with species such as SiO \citep{lee20,towner23}. More recently, in a few cases, jets have been detected in radio synchrotron emission \citep{anglada18,anton19}. However, no tracer is complete, and because of very high extinction, the deeply embedded regions, particularly the outflows from class 0/I protostars, must be studied at higher angular resolutions at longer wavelengths. While (sub)millimeter data from facilities such as ALMA have enabled detailed imaging of cold molecular material \citep{arce13, hsieh23}, mid-infrared (MIR) continuum and spectroscopic imaging of warmer atomic and molecular material is essential for a detailed characterization of the outflow processes. The James Webb Space Telescope (JWST) \citep{gardner23} with its MIR capabilities can identify and spatially resolve multiple flow components hidden at shorter wavelengths, providing a unique window into the deeply embedded protostellar phase.

\begin{figure*} [h!]
\begin{center}
 \sidecaption
\includegraphics[width=0.69\linewidth]{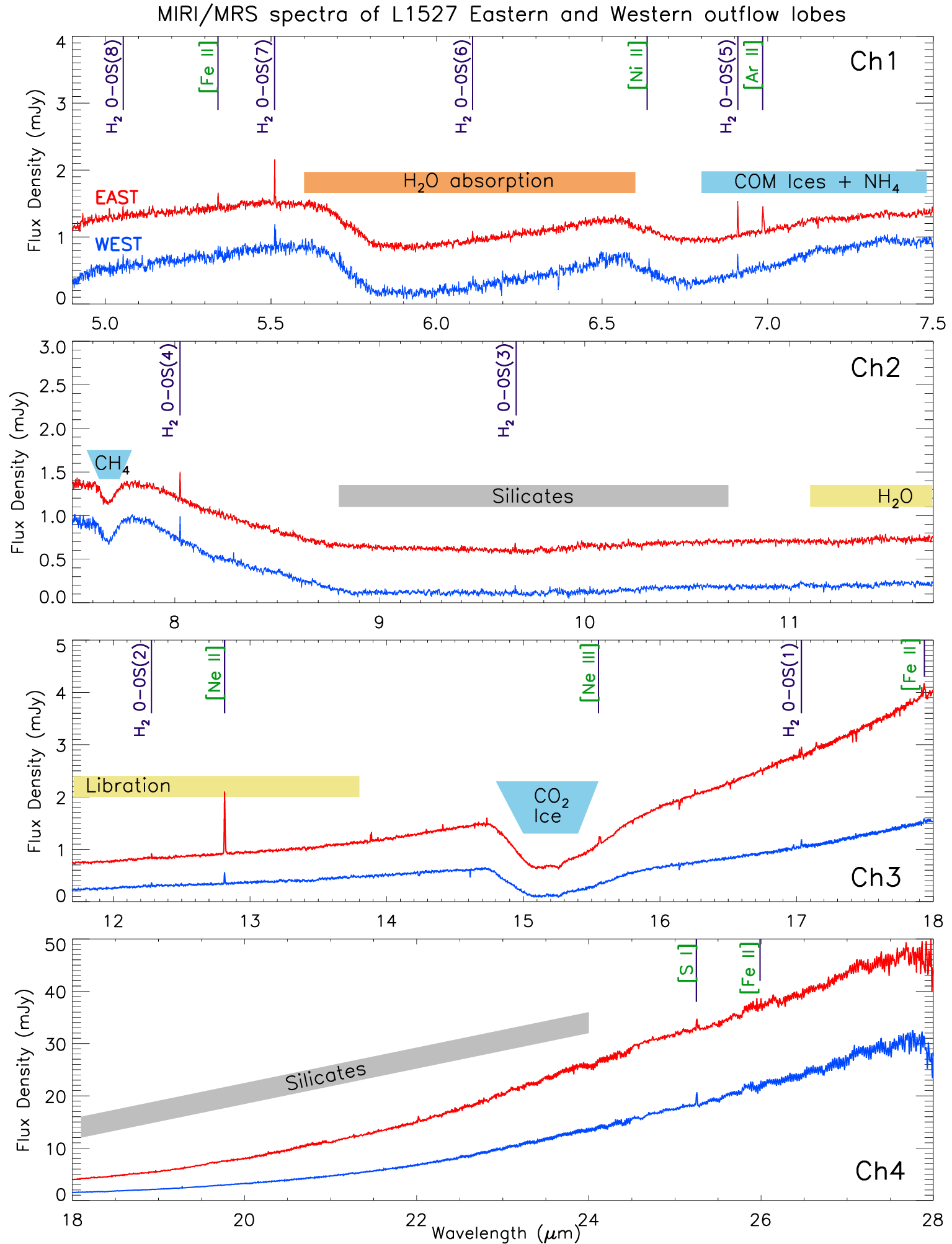}
 \caption{Detailed MIRI/MRS spectra extracted at the aperture positions shown in Fig.~\ref{fig1}. The four different panels correspond to the spectra from the four MIRI channels. Various molecular and ionized emission lines are identified and labeled. Major solid-state absorption features from ices and silicates are indicated by wide colored bands. The red spectra are scaled with increased offsets to avoid overlap with the blue spectra.}
\label{fig2}
\end{center}
\end{figure*}

The object of our study is L1527 IRS (also known as IRAS\,04368+2557; hereafter L1527), a low-mass protostar located in the L1527 dark cloud \citep{lynds62} in the Taurus star-forming region. It is estimated to lie at a distance of $140\pm1$\,pc by the most recent studies using Gaia data release 2 (DR2) and very long baseline interferometry (VLBI) data \citep{luhman18,galli19}. The protostar has been observed extensively from the NIR to centimeter wavelengths. The systemic velocity of the region in the Local-Standard-of-Rest (LSR) frame was estimated to be V$_{\rm{sys,LSR}} \approx 5.9$\,\kms \citep{ohashi97,tobin11}. The earliest evidence of outflow in L1527 was identified with (sub)millimeter observations in CO and HCO$^{+}$ emissions \citep{bontemps96,tamura96,zhou96,hoger98}. Their outflow maps revealed a redshifted and blueshifted emission oriented in the east-west direction in the plane of sky. A detailed picture of the outflow emission in scattered light, extending $\sim$20,000\,au, was presented by \citet{tobin08,tobin10} using the Spitzer Space Telescope and Gemini telescope. During Cycle 1 observations, JWST released\footnote{(Release ID 2022-055, PI: K. Pontoppidan) https://webbtelescope.org/contents/news-releases/2022/news-2022-055} high-resolution NIR broadband imaging of the outflow, revealing an intricate map of its dust layers and shocked material. Using ALMA data, \citet{vanthoff23} very recently identified a blueshifted unresolved SiO emission on the western lobe, indicating a tentative jet.

L1527 is often classified as a borderline class 0/I protostar due its low bolometric temperature ($\sim$40\,K) as well as its large and massive envelope ($\sim$2000\,au and 0.9 $\msun$) \citep{jorgensen07, tobin13, ohashi23}. It is one of the first protostars in its earliest evolutionary stage toward which a Keplerian rotating disk was established \citep{tobin12}. The disk has been viewed to be edge-on, and its inclination angle was estimated to be $i\approx75-85^{\circ}$ \citep{tobin08,ohashi23}. High-resolution sensitive observations of the disk have shown it to be warped and asymmetric, extending up to 100\,au \citep{sheehan22,vanthoff23}. A recent analysis of L1527's spectral energy distribution (SED) has estimated its bolometric luminosity to be in the range of $1.6-2.0\,\lsun$ and a protostar mass of about $0.2-0.4\,\msun$ \citep{aso17,karska18,ohashi23}. Using VLA observations at 7\,mm, \citet{loinard02} directly imaged the disk of L1527 and found an unresolved blob 25\,au from the central protostar, suspecting it to be a binary companion. However, many recent high-sensitivity observations with ALMA and VLA have not been able to confirm binarity \citep[e.g.,][]{nakatani20,sheehan22}. Although L1527 has been studied at several wavelengths -- revealing an edge-on disk, large envelope mass, and extended outﬂow cavities -- it has not been observed to have any clear collimated HV jet typical of class 0 protostars. This raises the question of whether it is steadily accreting in its current state and if definitive evidence of a jet exists at all. Therefore, L1527 is an ideal candidate for probing warm molecular and atomic jet emission using JWST MIR spectroscopy. Additionally, it is thought that the collimation and strength of jets decline with protostar age as the mass accretion rate decreases \citep[e.g.,][]{bontemps96, bally16}. Hence, discerning L1527's outflow properties will be key to understanding its evolutionary stage.

In this study, we present new results from JWST MIRI/MRS spectroscopic observations of L1527 covering the inner central region of 500\,au. In Section~\ref{obs}, we briefly describe the instrument and provide observational details. In Section~\ref{result}, we present the results of outflow spectra and spatial maps of different emission lines. In Section~\ref{temp}, we present the physical properties (temperature and column density) derived from excitation diagram analysis of H$_2$ lines and their comparison with shock models. In Section~\ref{discuss}, we discuss the results and presence of a jet. Finally, in Section~\ref{summary} we present the summary and concluding remarks.

\section{Observations}
\label{obs}

L1527 observations with JWST were obtained with the Mid-Infrared Instrument/Medium Resolution Spectrometer \citep[MIRI/MRS;][]{rieke15,wright23,argyriou23}. The observations were part of a larger program called ``JWST Observations of Young protoStars (JOYS)'' \citep[PID1290;][]{dishoeck25}. The JOYS program observes a total of about 20 low- to high-mass star-forming cores. The MIRI/MRS instrument does simultaneous spatial and spectral observations between wavelengths of $5\,\mu$m to $28\,\mu$m, covering a field of view of up to $\sim$7$''\times 7''$ with a spectral resolving power of $R\approx3500$ to 1500. The instrument has four IFU channels with three grating sub-band settings (short, medium, and long). A full spectrum coverage requires exposure in each of the three sub-bands, resulting in 12 spectral segments. MIRI/MRS achieves a point spread function (PSF) full width at half maximum (FWHM) of $\sim$0.2$''$ at $5\,\mu$m to $\sim$1$''$ at $28\,\mu$m, corresponding to a physical scale of $\sim$30\,au to $\sim$130\,au at a distance of 140\,pc, thereby providing coverage to the innermost region of L1527. 

Science observations were carried out on March 4, 2023, as part of Cycle 1 JOYS guaranteed time observations. The target position was centered on coordinates 4$^{\rm{h}}39^{\rm{m}}53^{\rm{s}}.955$ +26${^\circ}$03\arcmin09\farcs56 (J2000). This pointing was slightly offset from the protostar center (4$^{\rm{h}}39^{\rm{m}}53^{\rm{s}}.87$ +26${^\circ}$03\arcmin09\farcs58) to cover more of the brighter eastern outflow lobe while still observing part of the western lobe. The data were acquired using FAST1 readout mode with 180 groups of one integration per sub-band. The integration time per sub-band was 1000\,s, which resulted in a total integration time of 3000\,s to obtain the full spectrum. A two-point dither pattern was implemented to cover the extended morphology of the target. Additionally, a dedicated background field was observed with no dithering and with the same target integration time.

The data were reduced from Stage 0 uncalibrated raw files (.uncal) with the JWST pipeline v.1.12.5 \citep{bushouse23}, using corresponding the Calibration Reference Data System (CRDS) context file \texttt{jwst1174.pmap}. During Stage 1 processing, the raw files were processed for detector-level corrections using \texttt{Detector1 Pipeline} default settings. This included dark current removal, bad pixel flagging, nonlinearity correction, ramp jump corrections, and readout noise removal, resulting in images with count rates per second. During Stage 2 processing, the images were calibrated using the \texttt{Spec2 Pipeline} settings. Here, the data underwent WCS addition, flat-field correction, background subtraction, flux calibration, and fringe correction including residual fringe removal \citep{crouzet25}. An additional bad-pixel removal routine was applied for outlier detection using the Vortex Image Processing (VIP) package \citep{christiaens23}. Absolute flux calibration was performed from observations of spectrophotometric standards to convert the detector image to units of megaJanskys per steradian \citep{argyriou23}. Wavelength calibration was applied using MIRI/MRS reference files derived from lab measurements and observations of line emission from astrophysical standards. The wavelength scale was additionally corrected from the observatory reference frame to the barycentric reference frame. Finally, for Stage 3 processing, the \texttt{Spec3 Pipeline} step was applied, which included spectral extraction and cube creation. In total, 12 data cubes were created, corresponding to each of the MIRI/MRS channel and sub-bands. The spectra of all 12 sub-bands were stitched together to enable a single analysis of the full MIRI/MRS wavelength range.

Since the MIRI spectral cubes were calibrated in the barycentric reference frame, we applied a constant barycentric-to-LSR velocity correction of -10.1 \kms, appropriate for L1527 coordinates and the observation time. In addition, the velocities were shifted to the rest frame by correcting for the systemic velocity of +5.9 \kms.

\begin{figure*}[h]
\begin{center}
\includegraphics[width=1.0\linewidth]{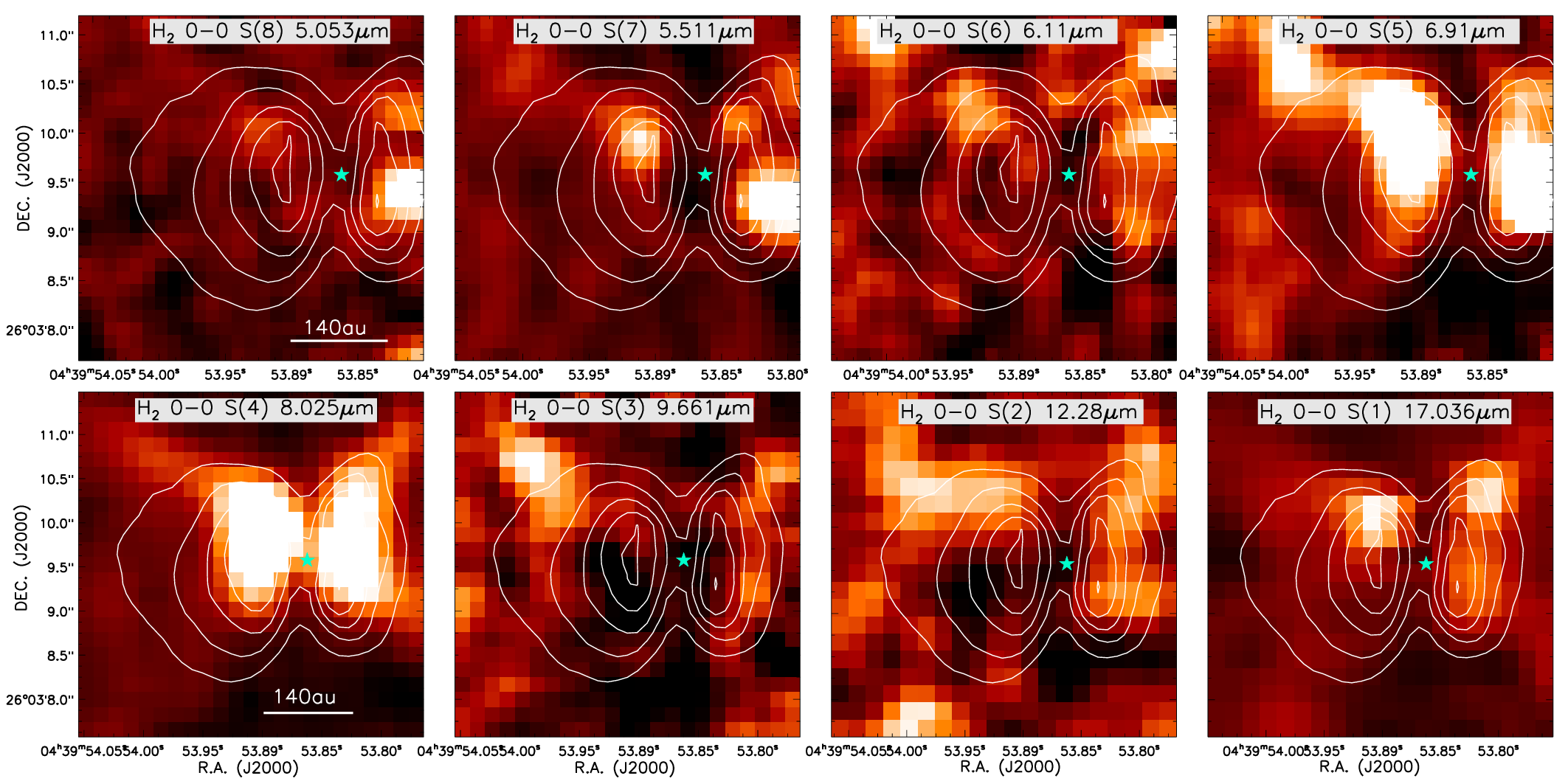}
 \caption{Line-integrated intensity (moment 0) maps of molecular H$_{2}$ 0-0 rotational transitions from S(8) to S(1) covering $5-28\,\mu$m range. The upper-state excitation energies for H$_{2}$ decrease from S(8) to S(1) transitions (see Table 1). The maps were created after subtracting the local continuum. The brightest regions correspond to the strongest H$_{2}$ emission. The white contours represent the $5.6\,\mu$m broadband emission. The position of the central star is marked with a star.}
\label{fig3}
\end{center}
\end{figure*}

\section{Results}
\label{result}

\subsection{General MIR imaging and spectral features}

In Fig.~\ref{fig1}, we present a MIR three-color map of the L1527 region, taken using MIRI broadband filters in $5.6\,\mu$m, $7.7\,\mu$m, and $12.8\,\mu$m (PID 1798: J. Tobin and PID 2729: K. Pontoppidan). This map, covering a large field of view, provides improved understanding of the outflow structure in the MIR compared to previous observations. The outflow and cavity walls are clearly seen as a bright bipolar structure extending to about 6000\,au in each direction. Several prominent nebulous filaments, consisting of bow shocks, knots, and reverse shocks that were unresolved in previous Spitzer MIR images \citep{tobin08}, are revealed in this new MIRI map. Some of the bright emission near the bow shocks may be associated with UV irradiation. Many cavities appear near the vicinity of the bow shocks where gas and dust are swept away. The outflow is launched at a wide angle from within the inner 200\,au region as previously observed and modeled \citep{tobin08, tobin13}. Toward the central region, the outflow structure is bright and contains the bulk of the emission, whereas it is fainter toward the wings and rear regions. Overall, the outflow appears to be in the plane of sky with the eastern flow relatively brighter than the western flow. 

The right panel in Fig.~\ref{fig1} shows a zoomed in view of the 500\,au central region with brightness scaled to reveal the eastern and western outflow lobes. Both lobes exhibit similar symmetrical morphology, with the eastern lobe being relatively brighter. The lobes are bisected by a dark lane at the center, as expected from the edge-on disk. The outflow lobes show bright continuum emission, which likely includes scattered light from the inner disk. The position of the forming protostar from millimeter continuum is marked by a star. Two circular apertures in red and blue are shown positioned on the outflow lobes for MIRI/MRS spectral extraction. These apertures were placed close to the central protostar while covering outflow emission, better representing the region's properties. The aperture radius was set to $R_{a} = 0.033 \times \lambda + 0.106$, which gives $0\farcs27$ at $5\,\mu$m and $0\farcs76$ at $20\,\mu$m. This is twice the MIRI PSF FWHM size as characterized by \citet{law23}.

%Describe more about continuum images with scattered light

The bottom panel in Fig.~\ref{fig1} shows the full distribution of the MIRI/MRS spectra extracted at the aperture positions. The color-coding of the spectra in red and blue correspond to the eastern and western lobes, respectively. Fig.~\ref{fig2} provides a more detailed view, separating the spectra into four panels corresponding to the four MIRI channels (Ch1, Ch2, Ch3, and Ch4), each observed with three grating settings. In general, the spectra increase in flux density with wavelength, which is typical for embedded protostars. However, several positions show strong absorption features characteristic of ices and silicates \citep{pont08, oberg11,yang22}. The ice features comprise both simple (H$_{2}$O, CO$_{2}$, and CH$_{4}$) and complex organic molecules (COM (CH$_{3}$OH, C$_{2}$H$_{5}$OH); \citep{rocha24,gelder24}). Strongest absorption appears particularly between $\sim$5 and $\sim11.5\,\mu$m. We also see sharp dips at a few positions around 7.7 and $15\,\mu$m due to CH$_{4}$ and CO$_{2}$ ice \citep{yang22}. These solid-state features are indicated by various colored bands in Fig.~\ref{fig2}.

At longer wavelengths, the increase in flux density is steep, as expected from the warm dust emission. Overall, the eastern and western lobes show a similar SED without any significant difference. Given the high sensitivity of JWST even with short integration times, we detect several spectral lines of molecular, atomic, and ionized emission toward both lobes. Pure rotational H$_{2}$ $\nu =$(0-0) transitions were detected covering the entire MIRI wavelength range. Forbidden atomic and ionized emission lines of \NiII, \ArII, \NeII, \NeIII, \SI, and \FeII\ were also detected. The spectra from the eastern region showed stronger intensities as well as more ionized emission lines, which are discussed in Section~\ref{alines}.

\begin{figure*}[h]
\begin{center}
\includegraphics[width=1.0\linewidth]{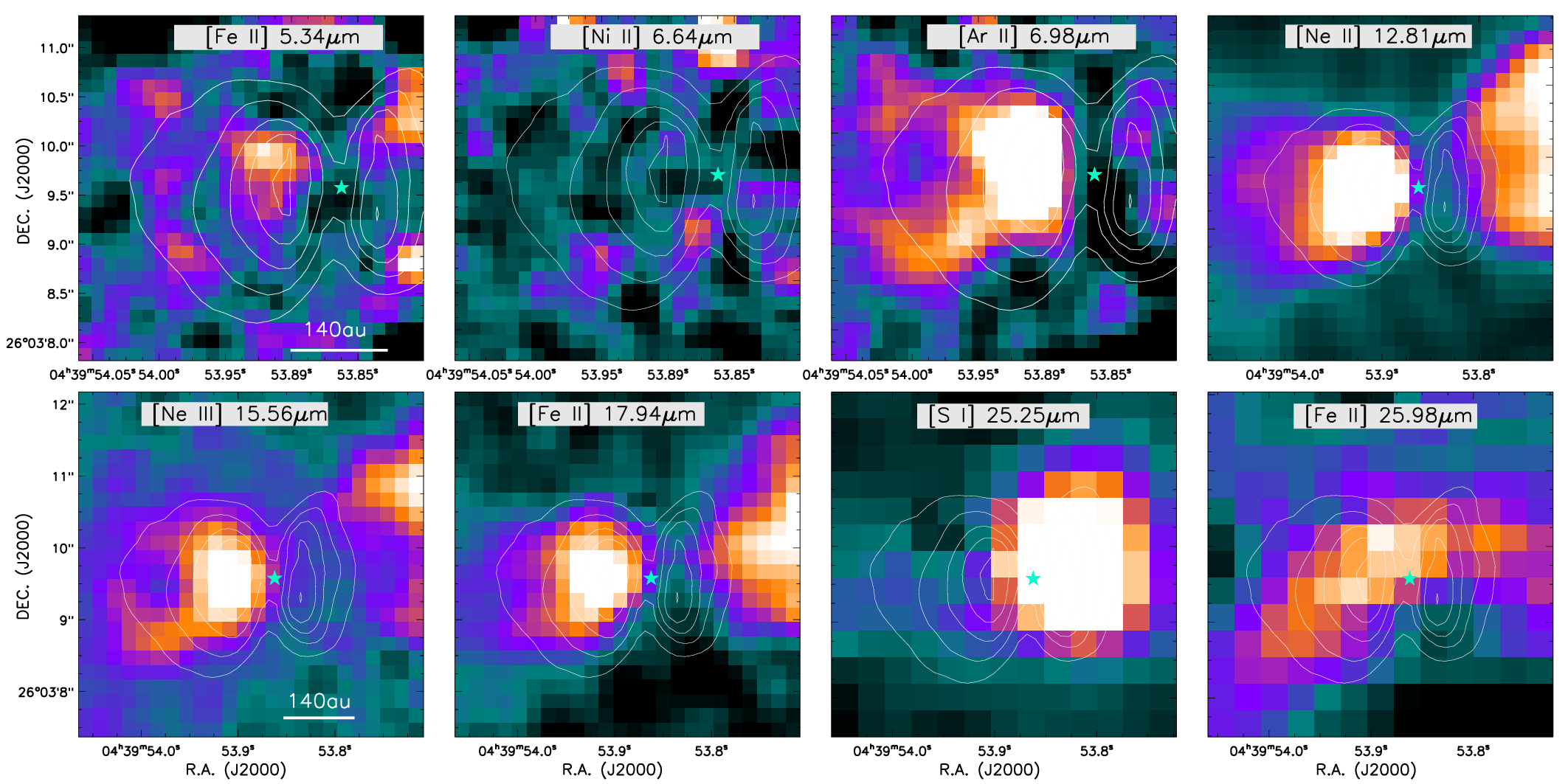}
 \caption{Line-integrated intensity (moment 0) maps of atomic and ionized emission lines in L1527. The maps cover all the lines detected in the MIRI spectral range and were created after subtracting the local continuum. The brightest white-orange regions correspond to the strongest line emission in each of the species. The white contours represent the $5.6\,\mu$m broadband emission. The position of the central protostar is marked with a star.}
\label{fig4}
\end{center}
\end{figure*}

\subsection{Molecular hydrogen emission}
\label{lines}

The MIRI/MRS spectral data cubes allow us to get a better understanding of the underlying outflow structures in L1527. Molecular hydrogen (H$_2$) is an excellent tracer of the bulk outflowing gas. These lines are primarily excited by shocks, with additional contributions from UV irradiation. Shock excitation exhibits a range of characteristics. Traditionally, $J$-type shocks are fast and dissociative, while $C$-type shocks are slower and largely non-dissociative \citep{hollen89,froebrich03}. However, the distinction between $J$- and $C$-type shocks depends not only on the shock velocity but also on the magnetic field strength and ionization fraction. In particular, slow $J$-type shocks with limited dissociation are possible under moderate magnetic fields \citep{kristensen23}. The H$_2$ emission can thus reflect a mixture of different shock conditions. The H$_2$ emission lines in the MIR are primarily due to rotational and rovibrational transitions \citep{caratti24, tycho24}. In our observation we detect the optically thin pure rotational $\nu =$(0-0) transitions from S(1) to S(8) shown in Fig.~\ref{fig2}. The S(0) line is not detected due to poor MIRI/MRS sensitivity beyond $27\,\mu$m. We find that the S(7), S(5), and the S(4) lines are the brightest in the spectra. Table~\ref{tabA1} provides the list of all detected H$_2$ lines with their properties.

In Fig.~\ref{fig3}, we present the continuum-subtracted H$_2$ line-integrated intensity (moment 0) maps with the field of view covering the inner 500\,au region. The general H$_2$ emission follows a bipolar morphology tracing the outflow structures. Transitions S(8), S(6), S(3), S(2), and S(1) are weak throughout the region, with detectable emission at only a few positions. Most of this is due to absorption from the ices and dust. The S(1) line appearing weak compared to higher excitation lines is particularly unusual and suggests that the emission is dominated by hot gas, pointing to excitation by shocks. The S(7) transition shows moderately detectable emission in both the lobes.

Transitions S(5) and S(4) have the strongest emission closely matching the morphology seen in the broadband images. The emission originates near the base of the outflow lobes and expands with a wide opening angle, outlining a symmetrical parabolic structure on both sides. The boundaries of the outflow cavity walls are clearly defined in these transitions and extend radially beyond the 200\,au region. The observed alignment likely reflects emission arising along the cavity walls, either from shocks or UV irradiation. While UV heating can affect the gas temperature, UV pumping primarily excites the vibrational levels of H$_{2}$ and has little impact on pure rotational transitions \citep[e.g.,][]{sternberg89}. Scattered H$_2$ emission is also expected to be negligible in the observations. While dust grains can scatter both continuum and line photons, dust scattering efficiencies at MIR wavelengths are generally low and are most significant in the NIR \citep[e.g.,][]{delab24}. Moreover, the line-to-continuum ratio in the observations is relatively low, so even if H$_2$ photons are scattered, their contribution to the observed emission is weak. Although scattering can be more important in some edge-on disks, such as HH\,30 \citep[e.g.,][]{tazaki25}, the observed morphology and brightness in L1527 indicate that direct H$_{2}$ line emission from the outflow cavities dominates. H$_{2}$ emission originating near the inner disk could in principle be scattered, but its expected contribution is also minor. Further discussion on continuum-scattered light is presented in \citet{tobin08,tobin10}.

No collimated features appear in any of the H$_{2}$ maps that might otherwise be associated with a jet. This is consistent with previous longer-wavelength molecular observations of wide outflow morphology \citep{yildiz15, vanthoff23}. The majority of the H$_2$ emission appears to be associated with a single morphological component of the outflow, most likely the outflow cavity itself. We observe no variation in the extent of the outflow opening angle with upper energy levels between each transitions. This behavior contrasts with that observed in TMC\,1-E \citep{tycho24}. The similar spatial extent of all the H$_2$ emission suggests a common origin within the outflow, although multiple excitation conditions may coexist in this region.

The overall excitation of the H$_2$ emission is most likely driven by shocks and shock-generated UV radiation, while direct stellar UV irradiation is unlikely to contribute significantly due to multiple constraints. Given that L1527 is a deeply embedded, low-mass protostar ($\sim0.3\,\msun$), the intrinsic stellar UV field is weak and can be heavily attenuated by the dense outflow and envelope ($\sim1.0\,\msun$), making direct UV irradiation of H$_{2}$ negligible. Moreover, no nearby massive star exists in the surrounding environment that could provide an external UV field. While some UV photons may preferentially escape along the outflow cavity, they would rapidly dilute long before reaching the spatial scales of the outflow ($>5000$\,au). The outflow morphology shows multiple discrete knots and bow-shaped structures aligned with the jet/outflow axis, characteristic of local collisional excitation rather than irradiation by a distant, centrally located source. Further evidence against a UV-dominated environment is the absence of the $11.2\,\mu$m polycyclic aromatic hydrocarbon (PAH) emission, which is commonly observed in protostars with JWST \citep[e.g.,][]{beuther23}. These factors together imply that large-scale UV irradiation cannot dominate H$_2$ line emission. Shocks from the disk wind and jets play a significant role in H$_2$ emission. Comparison of the observed data with H$_2$ shock models \citep{kristensen23} demonstrates that moderate-velocity $J$-shocks with modest UV field values appear most consistent with our results (see Section~\ref{shockmodel}).

\begin{figure*}[ht]
\begin{center}
\includegraphics[width=1.01\linewidth]{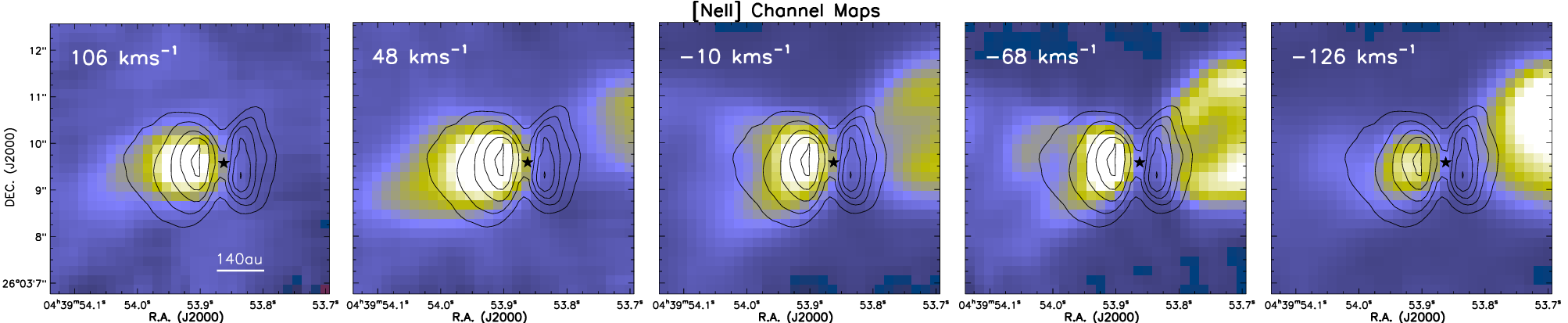}
 \caption{Channel map of \NeII\ $12.813\,\mu$m continuum-subtracted emission in L1527. The maps span the radial velocity (V$_{\rm{rest,LSR}}$) range between -126\,\kms to 106\,\kms. Extended emission tracing the outflow on both the lobes is shown. The black contours show the $5.6\,\mu$m broadband emission. Position of the central protostar is marked with a star.}
\label{fig5}
\end{center}
\end{figure*}

\subsection{Atomic and ionized Lines}
\label{alines}

Atomic and ionized species are useful for tracing various physical components within the outflow environment. Their emission reveals the underlying process involved in the excitation of these species \citep{neufeld09, beuther23, gieser23}. This is the first time several emission lines in the MIR have been observed in the L1527 outflow. As shown in Fig.~\ref{fig2}, we detect \NiII, \ArII, \NeII, \NeIII, \SI, and \FeII\ emission lines. Table~\ref{tabA2} provides the properties of the detected atomic and ionized lines.

In Fig.~\ref{fig4}, we present the continuum-subtracted, line-integrated intensity (moment 0) maps. The \FeII\ ($5.34\,\mu$m) and \NiII\ ($6.64\,\mu$m) maps have very faint emission without any significant morphological details. Noble gases of \ArII, \NeII, and \NeIII\ show the most prominent emission in the maps tracing the outflow structures. Their origin could be due to high-energy shocks from the outflow or photoionization by UV radiation. The \ArII\ map shows predominantly bright emission in the eastern lobe. This emission is distributed similarly to the H$_2$ S(5) emission, with both species excited at nearby central wavelengths. The bright \ArII\ emission could trace both the outflow cavity and the shocked material from the entrained ambient gas, reflecting increased ionization toward the eastern flow. On the other hand, there is almost no detection in the western lobe, suggesting that the shocked material has already been be cleared or lies farther from the central region. 

In the \NeII\ and \NeIII\ maps, we observe the full extent of the outflow material in both lobes. The emission exhibits a clear bipolar distribution, expanding outward along the direction of the outflow. On the eastern side, the emission remains closely aligned with the continuum and is more uniformly distributed along the outflow axis. However, on the western side the emission becomes significantly brighter only beyond a distance of approximately 150\,au from the central protostar, suggesting asymmetry in the excitation conditions or variations in the density of the medium. The presence of both \NeII\ and \NeIII\ emission indicates multiple ionization layers in the outflow, with \NeIII\ requiring higher energies for ionization. We discuss further details on the \NeII\ emission in Section~\ref{neii}.

The \FeII\ emission at both $17.94\,\mu$m and $25.98\,\mu$m shows outflowing shocked material similar to the \NeII\ morphology. However, the emission at $25.98\,\mu$m is much fainter and is affected by poorer resolution. The outflowing material seen here is not symmetric around both lobes. We see that the western ejection is much stronger and expands slightly northward compared to the H$_2$ emission. This is also not observed in the broadband images. Most of the \FeII\ emission traces the warm shocked gas. It likely arises from dissociative shocks at low excitation energy levels \citep{neufeld09}.

The \SI\ line is the only neutral atomic emission we detect in our observations, unlike other JOYS sources, which have detected \ClI\ and \FeI\ \citep[e.g., HH\,211;][]{caratti24}. The \SI\ map shows distinctively bright emission in the western lobe. The emission is not collimated and appears spatially extended. No significant \SI\ emission appears toward the eastern lobe, unlike the the ionized lines, which exhibit stronger emission there. The prominent \SI\ emission therefore most likely reflects different ionization conditions between the two outflow lobes, with the western lobe containing predominantly neutral gas. Such neutral conditions naturally favor excitation of the \SI\ line and are consistent with slow, partially dissociative $J$-shocks in dense gas \citep{hollen89, neufeld07}. In such environments, the enhanced \SI\ emission does not require a separate dust-destruction mechanism but instead is produced by shock-excited neutral gas in which the ionization fraction is considerably lower.

\begin{figure}[h]
\begin{center}
\includegraphics[width=0.9\linewidth]{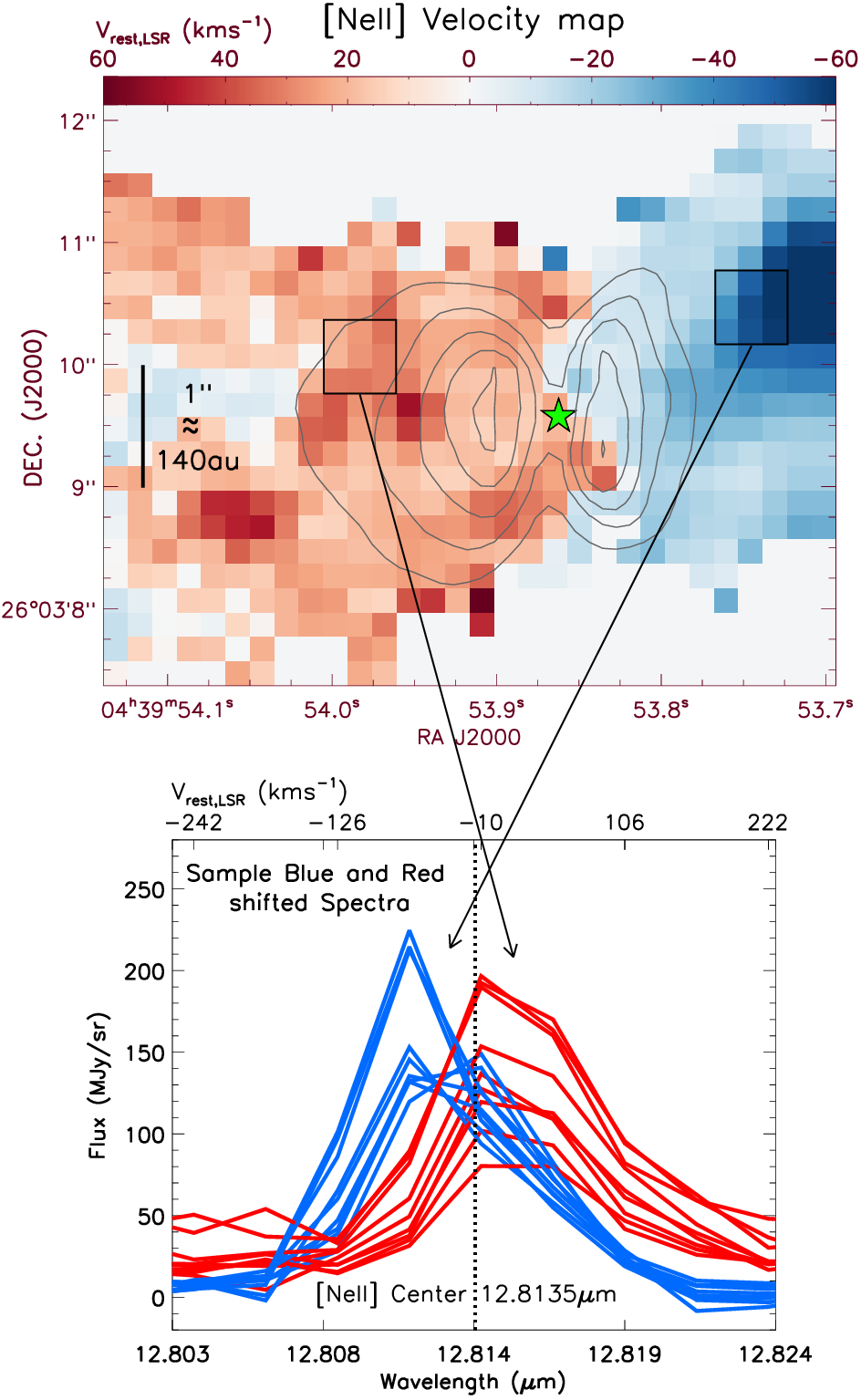}
 \caption{Top: \NeII\ radial-velocity (V$_{\rm{rest,LSR}}$) map of the L1527 region. The map was created by Gaussian fitting the \NeII\ line and estimating the Doppler-shifted velocity in each spaxel. The black contours show the $5.6\,\mu$m broadband emission. Bottom: Sample red- and blueshifted spectra for a box region of $3\times3$ pixels in the eastern and western regions. The central \NeII\ wavelength is shown by a vertical dotted line.}
\label{fig6}
\end{center}
\end{figure}

\subsection{\NeII\ line analysis}
\label{neii}

The MIR fine-structure line of \NeII\ at $12.81\,\mu$m is an excellent tracer of gas in outflows and disks. In outflows the line is thought to be excited by HV dissociative shocks of the ambient gas or UV/X-ray photoionization. In class II disks, it originates mainly from heating and ionization by stellar X-ray and UV radiation \citep{hollen89,pascucci09,gudel10}. The distance from the central protostar to the emission location generally determines its excitation condition. 

The \NeII\ line exhibits one of the strongest emissions among all the detected lines. As shown in the moment 0 map (Fig.~\ref{fig4}), the emission traces the characteristic structure of a bipolar outflow. However, the morphology of the emission is not spatially symmetric across both lobes. In the eastern region, most of the emission is focused near the outflow lobes. In contrast, in the western side, the emission appears fainter near the protostar but becomes significantly stronger beyond a distance of $\sim$150\,au. Additionally, the emission in the western side shows a slightly northward flow, consistent with the morphology observed in the \FeII\ maps.

In Fig.~\ref{fig5}, we present the \NeII\ continuum-subtracted channel maps at different radial velocities (V$_{\rm{rest,LSR}}$ between -126 to 106\,\kms). Each panel corresponds to the minimum spectral interval ($\sim 0.0025\,\mu$m) of the MIRI data cube. We see predominantly bright emission on the outflow lobe across all velocity ranges in the eastern region. This emission is very close to the protostar and most likely traces the outflow wind. In addition, we observe faint extended emission nearby that could arise from the outflow cavity walls.

In the western side, we see clear bright emission at blueshifted velocities. As we move from redshifted to blueshifted velocities, the strength of the emission increases progressively and highlights the different sections in the \NeII\ emission. Most notably, the emission at higher blueshifted velocities appears brighter farther away from the protostar. At about 250\,au away from the protostar, we see a bright blob of shocked emission suggesting that the dense ambient gas has been ionized by HV flows. Such types of emission is predicted to arise from $J$-shocks with high-shock velocities \citep{hollen89}. It is also predicted that \NeII\ intensities increase with the presence of high-density material, which may correspond to our results. Overall, we see that the \NeII\ emission is traced across multiple regions from near to far from the protostar, indicating a predominantly shock-driven origin.

In Fig.~\ref{fig6}, we present the radial velocity map derived from the \NeII\ emission lines. To obtain this map, we performed 1D Gaussian fits to each pixel in the data cube and estimated the radial velocity of the line, shifted to the rest frame in LSR. We set a minimum threshold of 10\,MJy\,sr$^{-1}$ for the Gaussian fits to ensure reliable velocities from high signal-to-noise (S/N) ratio lines. In the bottom panel of Fig.~\ref{fig6}, we present sample spectra for a box region of $3\times3$ pixels to show the distribution of the red- and blueshifted spectra with respect to the \NeII\ central wavelength. The resulting velocity map clearly shows the region to be distributed into two outflow components: the red- and blueshifted regions. Both velocity components appear to bifurcate near the location of the central protostar, supporting the interpretation that the outflow is oriented close to the plane of sky. This is consistent with previous estimates of inclination of $i\approx75-85^{\circ}$. There is also a slight presence of redshifted emission in the western lobe close to the protostar, which may be due to orientation or projection effects.

\begin{figure*}[h]
\begin{center}
\includegraphics[width=0.88\linewidth]{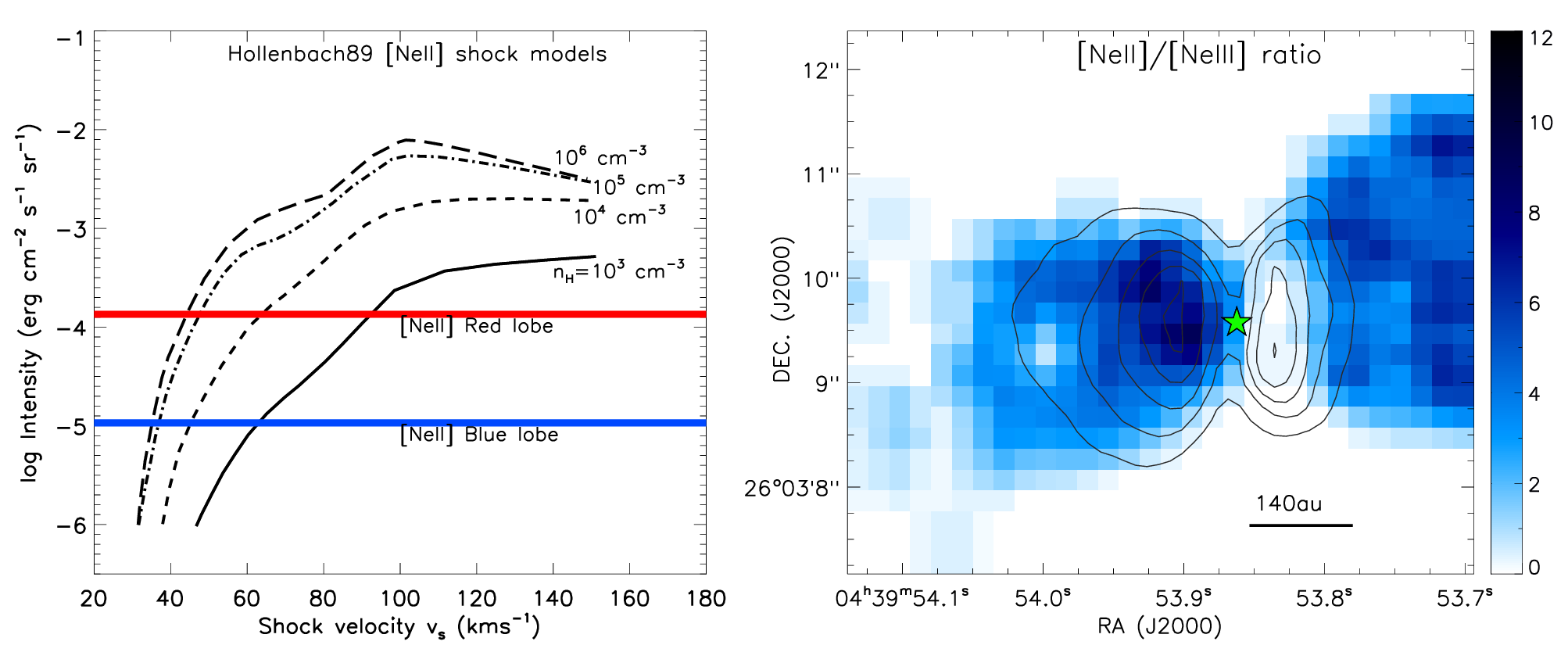}
 \caption{Left: \NeII\ emission compared to the \citet{hollen89} $J$-type shock models with a range of predicted intensities against pre-shock densities and shock velocities. Observed line intensities in red and blue were obtained at aperture positions in the eastern and western lobes (see Table~\ref{tabA2}). Right: \NeII$\mathbin{/}$\NeIII\ ratio map obtained from continuum-subtracted and extinction-corrected line intensities.}
\label{fig7}
\end{center}
\end{figure*}

\begin{figure*}[h]
\begin{center}
\includegraphics[width=0.9\linewidth]{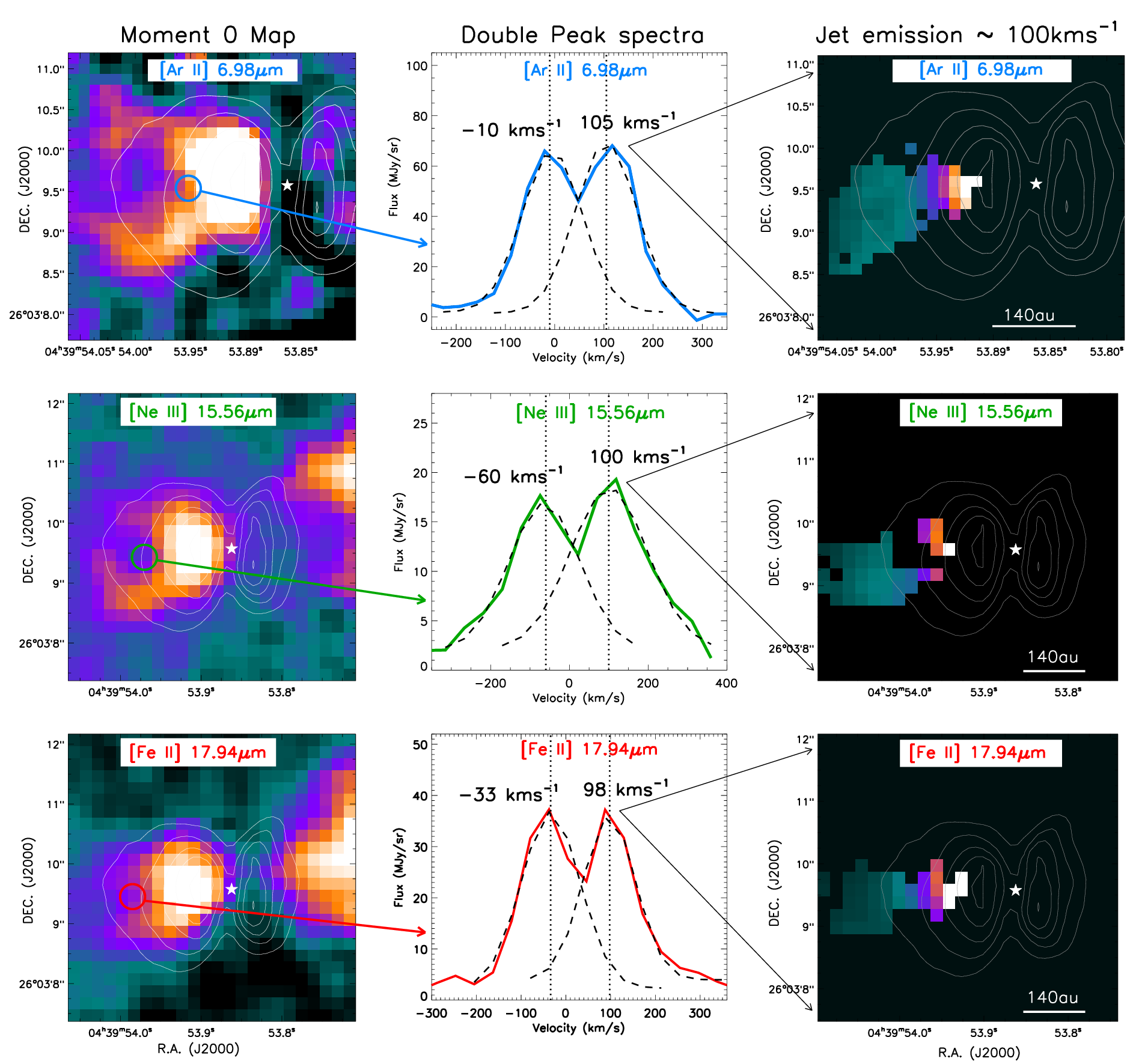}
 \caption{Double-peaked emission observed in the \ArII, \NeIII, and \FeII\ emission lines. Left column: Moment 0 map of the lines with aperture positions covering sample pixels with double-peaked emission. Middle column: Spectra of a double-peaked emission line in blue, green, and red for \ArII, \NeIII, and \FeII\,, respectively. 1D Gaussian fits to each velocity component are shown in dashed lines. The estimated radial velocity values of the low- and HV components are indicated. Right column: Distribution of the HV component associated with the jet emission. The brightest white-orange regions correspond to the strongest jet emission.}
\label{fig8}
\end{center}
\end{figure*}

The \NeII\ outflow material does not propagate strictly along the horizontal axis. Instead, the blueshifted region is observed to move northwest, consistent with the direction seen in other ionized emission lines. This type of morphology is apparent neither in the broadband images nor in the H$_{2}$ emission maps, suggesting that the molecular and ionized emission trace different layers of the outflow. The average velocity in the blue- and redshifted region is about $-24\pm8$\,\kms and $28\pm11$\,\kms, respectively. The redshifted region appears uniformly distributed without any significant velocity gradient along the flow. In contrast, the blueshifted region shows a sharp increase in velocity with distance from the protostar, consistent with the flow structure observed in the channel maps (Fig.~\ref{fig5}). Additionally, a vertical velocity gradient is evident in the blueshifted region, suggesting the presence of emission from multiple layers. The lower blueshifted velocities at the outer edges likely trace the outflow cavity wall, whereas the central brighter blueshifted feature appears to originate from shock-heated gas. 

The most interesting feature of the velocity map is the distribution of the redshifted and blueshifted velocities, which appear associated with the eastern and western regions, respectively. This completely contrasts a few earlier studies \citep{hoger98, tobin08}, which suggested that the eastern side is blueshifted and the western side is redshifted. \citet{yildiz15} conducted $^{12}$CO observations in the 3-2 and 6-5 transitions and reported that the $^{12}$CO(3-2) maps predominantly showed blueshifted velocities in the eastern side. However, in the $^{12}$CO(6-5) transition, they observed the opposite: blueshifted velocities on the western side. This mixed velocity distribution was attributed to the large opening angle of the outflow and its orientation nearly in the plane of sky.

High-resolution ALMA observations by \citet{oya15} using CS rotational lines, a dense gas tracer, provided a clearer kinematic picture. Their position-velocity diagram analysis suggested that the western side of the outflow is blueshifted, while the eastern side is redshifted, consistent with our findings in the \NeII\ velocity map. We also performed velocity analysis for other MIRI ionized lines and obtained a similar result. This result is somewhat counterintuitive in the context of scattered light continuum, where the eastern side appears brighter than the western side \citep{tobin08, tobin10}. Typically, dust scattering is brighter on the side of the disk tilted toward the observer. Recently, \citet{vanthoff23} showed brightening on the eastern side, suggesting that the extended disk is inclined toward the observer. However, given the flared and asymmetric nature of the disk, the inner disk may still not be tilted toward the observer. Thus, the discrepancy between the scattered light observations and the kinematic structure revealed by the \NeII\ maps suggests that the dust distribution may not be perfectly aligned with the true inclination of the disk. Furthermore, the MIRI/MRS data provide high-resolution spectral imaging of the innermost regions and are less susceptible to biases from large-scale wide morphological orientation. Therefore, our results taken together with the jet orientation (see Section~\ref{discuss}) likely represent the true orientation of the outflow, with the eastern side associated with redshifted velocities and the western side with blueshifted velocities.

\subsection{\NeII\ shock model and \NeII/\NeIII\ ratio map}
\label{neiishock}

\citet{hollen89} presented detailed $J$-type shock models that produce strong atomic and ionic emission through fast, dissociative heating at shock fronts in the absence of a significant magnetic field. In particular, the model predicts bright \NeII\ emission from HV shocks, which is sensitive to pre-shock density and shock velocity.

To assess whether $J$-shocks can account for the observed \NeII\ emission in L1527, we compared the models to our continuum-subtracted and extinction-corrected \NeII\ line intensities obtained at aperture positions along the eastern and western outflow lobes. The left panel in Fig.~\ref{fig7} shows the solid red and blue lines plotted against pre-shock density and shock velocity. The models that best reproduce the measured intensities correspond to shock velocities of approximately 35 to 45 \kms at pre-shock volume densities of $\sim10^{4}-10^{5}$\,cm$^{-3}$, which is typically expected in the dense envelope of L1527. This agreement supports the interpretation that moderate-velocity dissociative $J$-type shocks operate in the outflow, producing sufficient ionization to power the observed \NeII\ emission.

The \NeII/\NeIII\ line ratio is widely used to diagnose the ionization state and hardness of the radiation or shock environment \citep{gudel10,shang10}. Since \NeII\ (ionization potential 21.56\,eV) traces singly ionized gas, while \NeIII\ (41.96\,eV) requires much higher-energy photons or strong collisional ionization, their ratio is sensitive to both the ionizing spectrum and the electron temperature and/or density in the emitting gas. Low ratios ($<$1) are typically associated with highly ionized environments such as fast dissociative shocks ($>80$\,\kms) or strong FUV/EUV radiation fields, whereas high ratios ($>$2) indicate predominantly singly ionized gas, due to either softer radiation, slower shocks, or efficient recombination in dense regions. This ratio therefore provides a valuable constraint on the physical conditions and excitation mechanism of the gas.

We constructed a spatially resolved \NeII/\NeIII\ ratio map using continuum-subtracted and extinction-corrected line intensities across the field. The right panel in Fig.~\ref{fig7} shows the image of the ratio map. The values range predominantly between $3-10$, indicating that \NeII\ emission clearly dominates over \NeIII\ throughout the outflow. Ratios in this range are consistent with moderately ionized, dissociative $J$-shocks where the ionization fraction is significant but not high enough to produce strong \NeIII\ emission, thereby supporting shock-driven excitation rather than a hard stellar UV/X-ray source \citep{hollen09}. Moreover, the spatial structure of the ratio map shows particularly enhanced values at a few knots and/or shocks positions, which may indicate high-density regions or contributions from soft ionization produced by shock-generated UV radiation. In the next Section~\ref{dpjet}, we identify a HV component in the ionized lines, indicating the presence of a fast jet. If \NeII/\NeIII\ is evaluated specifically for the HV component, the ratio decreases to values $\lesssim1$. This behavior arises because the HV component is barely detected in \NeII, while clearly present in \NeIII, indicating that the \NeIII\ ionization is predominantly associated with the fast jet (with velocities $>80$\,\kms). This implies different excitation regimes within a stratified jet-outflow structure, where fast, highly ionizing shocks in the jet produce \NeIII, while moderate, partially ionizing shocks and shock-generated UV fields in the cavity produce the dominant \NeII\ emission.

\subsection{Double-peaked lines: Jet detection}
\label{dpjet}

Double-peaked spectral lines in protostars are linked to complex geometries and kinematics. These lines typically arise from Doppler-shifted emission caused by fast moving flows with their components moving at different velocities. High-resolution observations of many protostars have shown that forbidden emission lines often exhibit two velocity components, namely the low-velocity (LV) and HV component; see the review by \citet{eisloffel00}. Such velocity separations can also arise in bow-shock structures, where a curved shock front produces a range of shock velocities and spatially separated emission regions \citep{hartigan87}. These interpretations have been supported by theoretical simulations with magnetohydrodynamic (MHD) jet models \citep{garcia01,shang02}. They have shown the presence of different velocity components through their position-velocity diagrams.

Closer inspection of the L1527 MIRI/MRS spectra reveals clear double-peaked emission in three different forbidden lines: \ArII, \NeIII, and \FeII. The \NeII\ line shows a very weak, unresolved profile limited to just a few pixels; it was therefore excluded. The double-peaked emission was detected only in the eastern region of the outflow. Fig.~\ref{fig8} shows the profiles and the morphology of the double-peaked spectra for the \ArII, \NeIII, and \FeII\ lines. In the first column, the moment 0 maps are presented with the sample aperture positions. In the middle column, we show the double-peaked spectra extracted at the sample aperture position. The spectra are color-coded in blue, green, and red for \ArII, \NeIII, and \FeII, respectively. The line profiles clearly show two separate velocity peaks associated with the LV and HV components. A 1D Gaussian fit was performed on each velocity component to estimate their peak radial velocity, shifted to the rest frame in the LSR. In the last column of Fig.~\ref{fig8}, we show the jet emission map associated with the HV component. The jet emission was isolated from the double-peaked line profile by extracting the HV component, thereby identifying the associated spatial region. This was achieved by integrating the line intensity over a velocity range of $80-150$\,\kms, with the additional criteria that the emission exhibited a Gaussian profile and a line width exceeding the spectral resolution. All three maps show an extended jet with its brightest emission closest to the protostar. 

The LV component observed in the double-peaked lines is distributed throughout the region and associated with the shocked gas in the general outflow or disk wind. The average radial velocity of the LV component is about $-30\,$\kms, which is notably blueshifted. However, we can confirm that in regions without double-peaked profiles, the average radial velocity is redshifted and is consistent with the \NeII\ velocity maps. Disk winds launched from a few to several au from the protostar are expected to have velocities of tens of kilometers per second \citep{pudritz07,shang07}. They are thought to shape the molecular outflow cavities by sweeping up the surrounding envelope material. Our observed LV component velocities are consistent with these predictions and support previous studies indicating that the L1527 outflow cavity is shaped by a slow disk wind entraining the ambient material \citep{tobin08,tobin12}.

The double-peaked HV component is present only in the eastern region of the outflow. The \ArII\ line shows the strongest emission among the three detected lines. The average radial velocity of the HV component is about $100\,$\kms. The emission is neither narrow nor collimated. Rather, it is extended in width and follows the direction of the outflow axis to about 500\,au. We see that the emission bends or curves toward the south in all three maps. The HV component is interpreted as evidence for the presence of a jet. Although the emission is not strictly collimated as in other class 0/I jets, its observed velocity ($\sim$100\,\kms), extended scale, and detection in high-ionization lines (\ArII, \NeIII) strongly suggest shock-excited jet activity, a strong characteristic of early-stage protostellar evolution. 

Considering the jet is almost in the plane of sky, we can correct the radial velocity for inclination. For a value of $i\approx75^{\circ}$, we obtain a true velocity of around $385\,$\kms. This value should be interpreted cautiously, as the disk is warped and the jet inclination varies significantly due to its precession and bending. Very high velocities ($>$200\,\kms) are predicted by magneto-centrifugal launching models when jets originate very close to the protostar. X-wind models \citep{shu07}, which launch material near the corotation radius, naturally produce such high velocities due to the strong gravitational potential in this region. However, disk-wind models \citep{ferreira97} can also generate comparably high velocities when launched from sub-au radii. Therefore, the high velocities observed in the jet likely indicate that its launching region lies very close to the protostar. Overall, the coexistence of both velocity components in multiple ionized lines suggests L1527 is composed of a stratified layer of gas with varying excitation conditions.

\begin{figure*}[h]
\begin{center}
\includegraphics[width=1.01\linewidth]{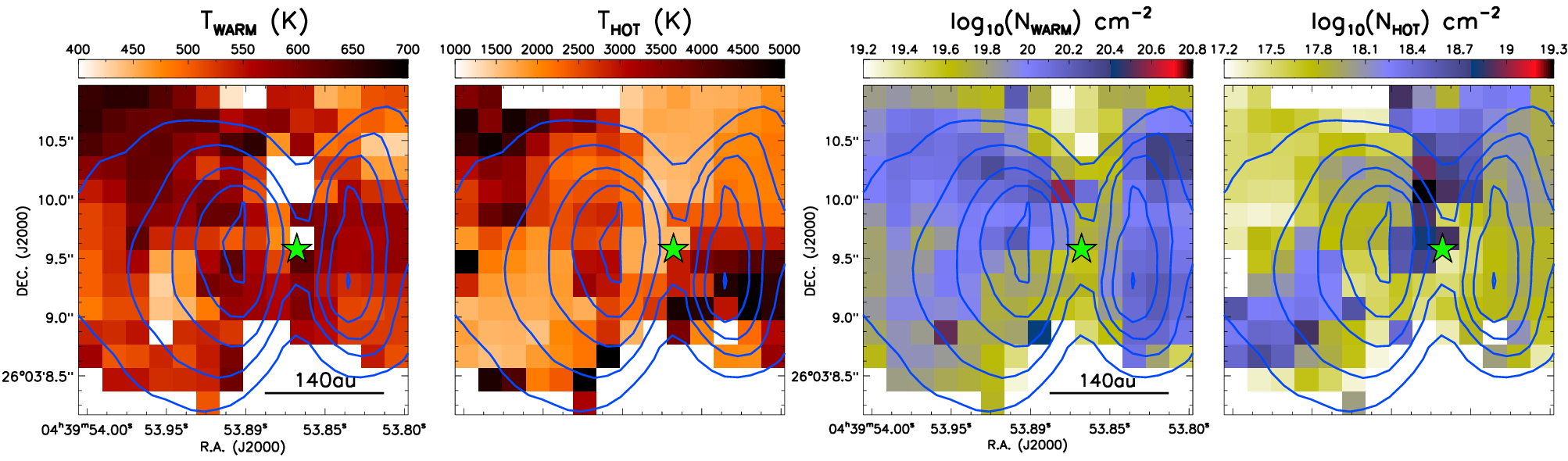}
 \caption{H$_{2}$ excitation diagram analysis resuts for L1527, obtained by applying the analysis to each spaxel across the entire region. Left: Temperature maps of the warm and hot gas components. Right: Corresponding column density maps. Color bars above each panel indicate the value ranges. The blue contours show the $5.6\,\mu$m broadband emission, and the central star is marked with a star.}
\label{fig09}
\end{center}
\end{figure*}

\section{Excitation diagram analysis} 
\subsection{Temperature and column density maps}
\label{temp}

Physical properties such as gas temperature $T_\mathrm{H_{2}}$ and column density $N$(H$_{2}$) can be inferred using excitation diagram analysis of H$_{2}$ rotational transitions. The observed extinction-corrected H$_{2}$ line-integrated intensities are converted into upper-state column densities normalized by its statistical weight ($N_\mathrm{u}$/$g_\mathrm{u}$). These are plotted in a logarithmic scale against their upper-state excitation energies ($E_\mathrm{u}$/$k_\mathrm{B}$). For gas in local thermal equilibrium (LTE), the gas excitation follows a Boltzmann distribution ($N_\mathrm{u}$/$g_\mathrm{u} \propto \exp(-E_\mathrm{u}$/$k_\mathrm{B})$), and the data points in the diagram can be fit with a straight line. The gas excitation temperature is derived as the reciprocal of the slope of the line, whereas the column density is proportional to the y-axis intercept. This analysis assumes that all H$_{2}$ transitions are optically thin, which is valid even for $N$(H$_{2}$)$>$10$^{23}$\,cm$^{-2}$ \citep{bitner08}. 

The MIRI/MRS spectral range covers several pure rotational H$_{2}$ transitions from the S(8) to S(1) state, which are detected in most pixels in the entire field of view. This allows us to carry out pixel-by-pixel excitation diagram analysis and to obtain a complete map of $T_\mathrm{H_{2}}$, and $N$(H$_{2}$). We first smoothed the spectral cubes from different MIRI channels to the lowest common resolution, as the angular resolution varies from 0\farcs2 to 1\farcs0 between all the MIRI channels. We smoothed the spectral cubes to a resolution of 0\farcs7, matching that of the S(1) transition, which is the lowest angular resolution in the detected H$_{2}$ lines. We also regridded the cubes to the same spatial grid as the H$_{2}$ S(1) data.
	
Next, we corrected for extinction along the line of sight. The most commonly used extinction correction in MIR comes from the \citet{mcclure09} and KP5 \citep{pont24} curves, with the latter introducing slightly stronger absorption from ices and silicates. Previous analysis of other JOYS sources adopted the \citet{mcclure09} extinction curve valid between a $K$-band magnitude of $A_K=$ 1 to 7\,mag. However, a recent study of ices in L1527 by \citet{slavic25} showed substantial ice and silicate absorption, indicating that the KP5 curve provides a more appropriate extinction correction, which we apply in our analysis. The optical depth at 9.7$\mu$m was earlier estimated to be around $\tau_{9.7}\sim 4$ \citep{slavic25}. We also measured the extinction by de-reddening the S(1) to S(8) lines in the excitation diagram analysis, which yielded $A_K = 7\,$mag (see Fig.~\ref{figb2}). Based on these results and previous extinction estimates from SED modeling \citep{tobin08,tobin10}, we adopted a value of $A_K = 7\,$mag, which is roughly equivalent to a visual extinction of $A_V\approx50$\,mag. We tested lower extinction values between $A_K = 4-6$\,mag and find that the intensity of the S(3) line decreases significantly and gets excluded from the fit. We also find that an H$_{2}$ ortho-to-para ratio of 3 provides the best fit to our observed line intensities, assuming the H$_{2}$ gas is in LTE.

The excitation diagram analysis was performed using the \texttt{pdrtpy}\footnote{The \texttt{pdrtpy} project, developed by Marc Pound and Mark Wolfire, is supported by the NASA Astrophysics Data Analysis Program grant 80NSSC19K0573; the JWST-ERS program ID 1288 through grants from STScI under NASA contract NAS5-03127; and the SOFIA C+ Legacy Project through NASA award \#208378 issued by USRA.} python package, which is commonly employed in the analysis of JOYS sources \citep{gieser23, caratti24, francis25}. Typically, the observed H$_{2}$ rotational transitions trace two temperature components, often parameterized as a "warm" and a "hot" component in excitation diagrams. The warm component ($T_\mathrm{warm}\approx200-1000$\,K) is traced by the lower excitation lines (e.g., S(0)-S(4), $E_{\rm up}\leq$ 4000\,K) and is commonly associated with slow, non-dissociative $C$-shocks or the cooler post-shock gas within $J$-shocks. The hot component ($T_\mathrm{hot}\approx2000-3000$\,K), traced by the higher excitation transitions (e.g., S(5) and above, $E_{\rm up}\geq$ 5000-10000\,K), arises mostly from energetic fast dissociative $J$-shocks where collisional heating dominates. However, recent shock models \citep[e.g.,][]{kristensen23} demonstrate that a single shock structure can produce both temperature components, where the hot gas represents the post-shock region and the warm gas traces the cooling layer of the same shock. Therefore, the presence of two temperature components does not necessarily indicate separate shocks but can instead reflect a thermally stratified region with distinct excitation conditions. In general, the warm component typically carries most of the H$_{2}$ luminosity and therefore remains a dominant tracer of the bulk outflowing gas, while the hot component highlights localized regions of strong shock activity.

At the initial step, the extinction-corrected H$_{2}$ transitions at each spaxel were fit by a 1D Gaussian, and the line intensities were computed from the integral of the Gaussian fit. We set an intensity threshold of 20\,MJy\,sr$^{-1}$ to avoid any spurious detection. In Fig.~\ref{figb1}, we show example spectra for spaxels associated with the apertures in the eastern and western lobes (see Fig.~\ref{fig1}). The spectra are plotted for different H$_{2}$ transitions, and their Gaussian fits are highlighted in red and blue for the eastern and western lobes, respectively. We see that the line intensities vary across different transitions depending on their strength. In general, the lines from all transitions have bona fide profiles for the Gaussian fits without any omission. 

Next, the line intensities from the spectra were converted to upper-state column densities and were then plotted against their excitation temperature, as described earlier. To obtain reliable estimates from the excitation diagram, fitting was performed only if five or more transitions were detected in a spaxel. Fig.~\ref{figb3} shows the excitation diagram for the example spaxels in the eastern and western lobes. The data points with error bars represent the upper-state column densities of the lines in each H$_{2}$ transition. They are distributed across the plot with overlays of red and blue fits for the warm and hot components, respectively. The total fit to the data points is shown as a solid green line.
The estimated warm temperatures in both the spaxels are similar, with an average of about $T_\mathrm{warm}\approx$570\,K. Whereas, the hot temperature varies considerably with values of $T_\mathrm{hot}\approx$2170\,K and 3340\,K for the eastern and western spaxels, respectively. The estimated warm and hot column densities in both spaxels average $N(\rm{H_{2}})_\mathrm{warm}\approx 16\times10^{19}\,\mathrm{cm}^{-2}$ and $N(\rm{H_{2}})_\mathrm{hot}\approx 9\times10^{17}\,\mathrm{cm}^{-2}$.

\begin{figure*}[h]
\begin{center}
\includegraphics[width=0.94\linewidth]{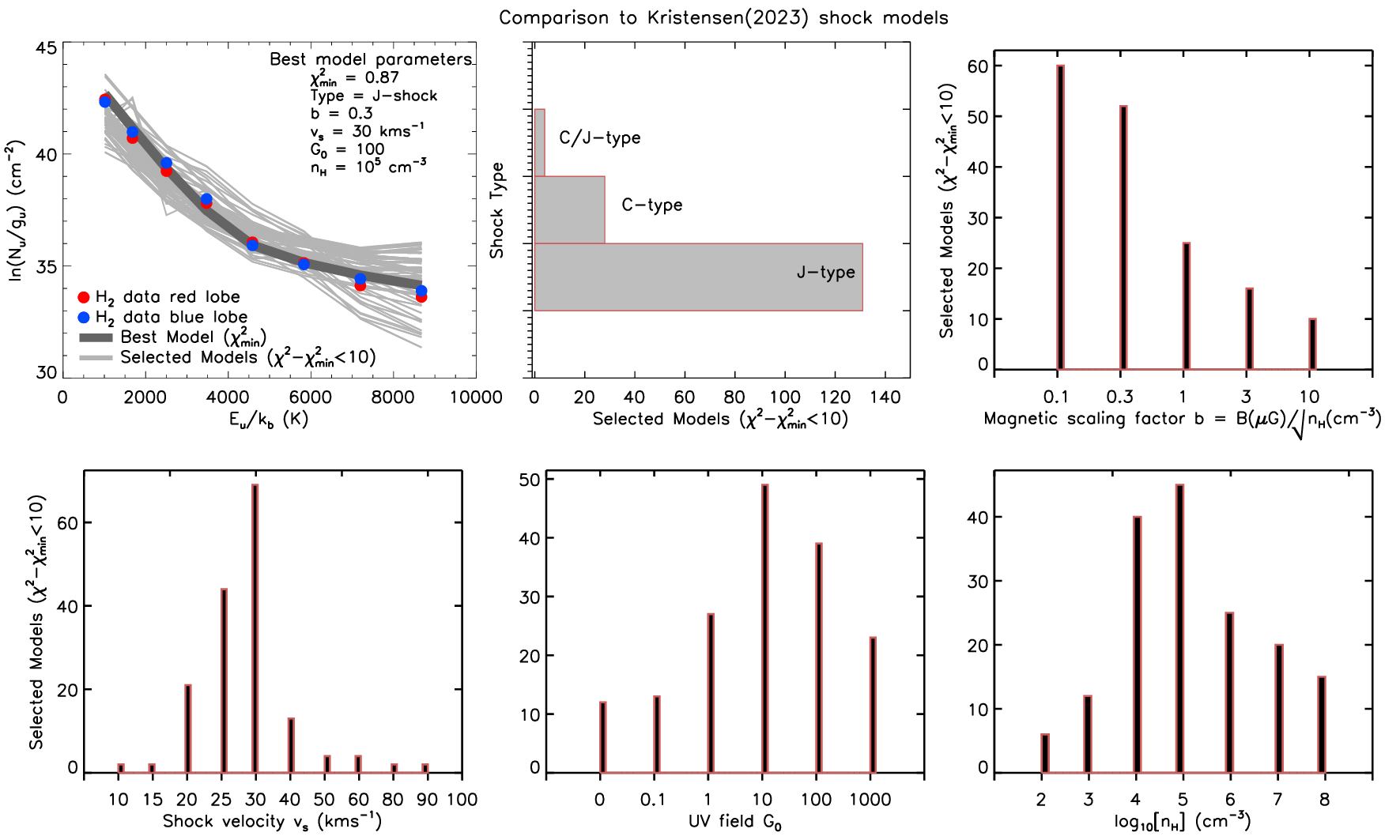}
 \caption{Top left: Extinction-corrected H$_{2}$ rotational lines vs. \citet{kristensen23} shock models. The best model with ${\chi^2_{\rm{min}}}$ is shown in solid dark gray lines with its parameters displayed. The selected models with ${\chi^2 - \chi^2_{\rm{min}}} < 10$ are overplotted with light gray lines. Subsequent panels show the histogram distribution of key model parameters (shock type, magnetic scaling parameter, shock velocity, external UV field, and pre-shock density) that determine the prevailing shock condition.}
\label{fig10}
\end{center}
\end{figure*}

Fig.~\ref{fig09} presents the full temperature and column density maps for both the warm and hot components. In the temperature maps, the warm component exhibits moderate temperature variations across the outflow region, with values ranging between $450-650$\,K. There is a noticeable gradient in the eastern lobe with temperatures increasing toward northeast. This gradient closely follows the spatial distribution of the H$_{2}$ emission, particularly resembling the brighter S(5) and S(4) transitions (see Fig.~\ref{fig3}). This suggests the presence of warm gas within the outflow cavity. The western lobe also has a gradient, though with temperatures increasing toward the southwest and with lesser resemblance to the H$_{2}$ emission. In the hot component map, temperature variations are substantial, ranging between $1500-5000$\,K. Both sides of the region exhibit sharp increases in temperature along the outflow structures, closely resembling the distribution of the H$_{2}$ emission. These temperature variations and their spatial resemblance to the outflow morphology suggest that gas heating is strongly linked to a common excitation mechanism within the outflow.

The column density maps exhibit a morphology broadly similar to the temperature maps. The warm component of the column density ranges from $N(\rm{H_{2}})_\mathrm{warm}\approx 2-70\times10^{19}\,\mathrm{cm}^{-2}$, whereas the hot component ranges from $N(\rm{H_{2}})_\mathrm{hot}\approx 2-100\times10^{17}\,\mathrm{cm}^{-2}$. The warm component is significantly denser, about two orders of magnitude higher than the hot component. In the warm component maps, regions of higher column densities show a general positive correlation with regions with higher temperatures. However, this is in contrast to the hot component maps, where regions of lower column density exhibit higher temperatures. These trends likely reflect the nature of shock heating in the outflow. 

In both the temperature and column density maps, we do not find any evidence of a jet based on the spatial variations or gradients at the position of the HV component (see Fig.~\ref{fig8}). This lack of jet temperature or density signature is consistent with the absence of a molecular jet in our H$_{2}$ maps. Overall, the gas temperature and column density maps exhibit a strong correlation with the outflow morphology, suggesting that the outflow cavity is composed of layers of warm and hot gas.

\subsection{Comparison to shock models}
\label{shockmodel} 

The H$_{2}$ emission from outflows in dense regions ($>10^5$\,cm$^{-3}$) is generally thought to be dominated by collisional excitation from shocks. However, UV-driven irradiation may still play a role if accretion-generated UV photons penetrate the base of the outflow cavity before being attenuated, or if fast dissociative shocks generate local UV fields that modify the shock structure and chemistry. The strong gradients in temperature, density, and ionization across shock fronts naturally produce multiple excitation components, with hotter gas populating higher-energy rotational states and warmer gas dominating lower levels. To constrain the mechanism of excitation in the outflow, we compared our observed H$_{2}$ emission with recent shock models that self-consistently treat different excitation processes.

We used the recent shock model grid from \citet{kristensen23}, based on the Paris-Durham code \citep{flower03,godard19}, which computes the dynamical, thermal, and chemical structure of steady, plane-parallel shocks over $\sim$14,000 models. The code determines pre-shock equilibrium conditions and includes self-consistent treatments of magnetic fields, UV irradiation, grain and PAH chemistry, and H$_{2}$ excitation and cooling. Depending on the shock speed and magnetic field strength, it solves for $C$, $J$, or $C/J$-type shocks. From each model, key outputs such as temperature, normalized upper-state column density, H$_{2}$-integrated intensities, shock width, and molecular abundances are provided. The comparison of JWST H$_2$ observations with shock grids has recently been studied in other outflows \citep[for e.g., see][]{gou25, navarro25}, demonstrating the utility of the models for constraining shock properties.

The \citet{kristensen23} shock grid explores six key input parameters: the pre-shock density $n_{\rm{H}}$ ($10^2-10^8$\,cm$^{-3}$), the magnetic field scaling parameter $b$ ($0.1-10$, where transverse magnetic field strength B($\mu$G) = $b\sqrt{n_{\rm{H}}}\,$), the shock velocity $v_s$ ($2-90$\,\kms, dependent on $b$), the external UV field strength $G_0$ $(0-10^3)$, the H$_{2}$ cosmic-ray ionization rate $\zeta_{\rm{H_2}}\, (10^{-17}-10^{-15}\,\rm{s}^{-1})$, and the PAH abundance $X(\rm{PAH})$ $(10^{-8}-10^{-6}$). The parameters $n_{\rm{H}}, G_0, \zeta_{\mathrm{H_2}}$, and $X(\rm{PAH})$ are sampled in factors of 10, while $b$ and $v_s$ vary non-uniformly.

We compared the observed extinction-corrected upper-state column densities from both outflows lobes (see Fig.~\ref{figb3}) with the corresponding values predicted by the models. We did not restrict the initial conditions and explored the full parameter space. For each model, the reduced $\chi^2$ was calculated to assess how well it reproduced the observed emission, and the model with minimum ${\chi^2_{\rm{min}}}$ was initially identified as the best fit. However, we found that several models can yield $\chi^2$ values close to the minimum while having differing parameter values. To account for this degeneracy and obtain robust constraints on the likely physical conditions of the shock, we considered all models with ${\chi^2 - \chi^2_{\rm{min}}} < 10$. This value was chosen to provide a practical compromise by excluding clearly discrepant models while allowing exploration of the full parameter degeneracies in the model grid. The selection criteria resulted in 156 models that closely matched the observed emission. Fig.~\ref{fig10} shows the best and selected model fits to the observed data.

We next explored the properties of the selected models to evaluate the most common and statistically significant parameter values. Histogram analysis of the key different parameters are shown in Fig.~\ref{fig10}. The analysis revealed that the majority of the selected models correspond to $J$-type shocks, indicating that these shocks are the dominant excitation mechanism in the observed regions. The magnetic field scaling factor $b$ is low, with values between 0.1 and 0.3, consistent with the absence of strong $C$-type shock signatures. The shock velocities were found to be generally moderate, around $30$\,\kms, while the pre-shock densities range from $n_{\rm{H}} = 10^4-10^5$\,cm$^{-3}$. The external UV radiation field $G_0$ was distributed mostly between 10 and 100, suggesting that UV irradiation contributes alongside collisional excitation. The $\zeta_{\rm{H_2}}$ and $X$(PAH) parameters have minimal impact, as shock modeling does not significantly depend on them \citep{kristensen23}.

These results highlight the significance of UV irradiation and suggest that, even if stellar UV is largely attenuated by the dense envelope, shock-generated UV remains a substantial contributor to the shock structure, especially its temperature. At velocities $\geq$30-50\,\kms, post-shock temperatures reach $\geq10^5$\,K, leading to H$_{2}$ dissociation and partial ionization. The hot post-shock gas then emits UV photons through both recombination and collisional excitation, creating a radiative precursor that enhances local photoionization. A recent study by \citet{navarro25} in HH\,46 similarly found that $J$-shocks accompanied by a local UV field dominate H$_2$ excitation. Therefore, moderate-velocity dissociative $J$-shocks can provide a natural in situ UV source while collisionally exciting both the warm and hot H$_{2}$ components that dominate the observed emission. The relatively weak magnetic field further suppresses the formation of extended $C$- type shocks. Overall, our results indicate that moderate-velocity $J$-type shocks, combined with a modest UV field, provide the best match to the observed H$_{2}$ emission in L1527. 

\section{Discussion}
\label{discuss}

JWST observations reveal a detailed hourglass-shaped bipolar outflow in the MIR broadband images (see Fig.~\ref{fig1}). The images exhibit multiple nebulous features consisting of warm and hot layers of gas and dust generated within the outflow cavity. Our spectral maps reveal that the emission includes molecular, ionized, and scattered light. Despite such detailed composition, we detect no trace of a large-scale collimated jet in the broadband images. This absence of a large jet signature is distinctive for L1527 given its evolutionary stage. In contrast, recent JWST observations of several other class 0/I protostars, such as HH\,211, HH\,46, TMC\,1, and IRAS\,16253-2429, have shown prominent jet emission. Their bulk of the emission is particularly traced in high-J H$_{2}$ and \FeII\ lines \citep{ray23,nisini24,tycho24,narang24}. We see that most of L1527's dominant emission comes from the wide-angle outflow across both molecular and ionized emission. However, the detection of double-peaked ionized lines with the HV component at velocities $\geq$100\,\kms provides compelling evidence for the presence of a jet. This jet signature spans less than 500\,au from the protostar and is fainter than the jets from other sources. In addition, compared to the large-scale outflow structures ($>$3000\,au), the jet is small and covers only a fraction of the region, thereby concealing its visibility in the broadband images.

Double-peaked emission lines with kinematics similar to L1527 have been observed in other protostellar outflows. For instance, HH\,34 presents a collimated jet showing two velocity components both in molecular and ionized lines \citep{nisini16}. Using VLT observations of H$_{2}$ and \FeII\ lines, \citet{garcia08} showed that the HH\,34 LV component is distributed at $\sim$15-50\,\kms surrounding a fast axial jet ($\sim$90-140\,\kms). Likewise, in DG\,Tau, \FeII\ and \NeIII\ double-peaked lines were observed with VLT, using SINFONI and X-shooter showing a similar velocity separation \citep{agra11,liu16}. Many other sources such as SVS\,13, HH\,211, and HH\,46 \citep{takami06,tabone17,garcia10} exhibit the same pattern. Such profiles are rarely seen in class 0 protostars due to their dense envelopes but are more notable in class II protostars due their moderate extinction. A unifying scenario in all these observations is that the LV component is associated with a slower, wider wind, while the HV component traces the faster jet, thereby pointing to a stratified structure in the outflow. 

The observed double-peaked lines can also be interpreted within the framework of shock models. The coexistence of the LV and HV components in the ionized lines suggests a stratified velocity structure within the predominantly $J$-type shock environment. Our comparison of models shows that moderate-velocity ($\sim30$\,\kms) $J$-type shocks illuminated by nonzero UV field best reproduce the observed emission. This suggests that the bulk of line excitation is driven by compact, dissociative shock fronts likely associated with the disk wind. The high \NeII/\NeIII\ across most of the emitting region indicates soft ionization overall, consistent with moderate-velocity shocks where UV photons generated in the shock front contribute to ionization. However, the clear detection of the HV component in \ArII, \NeIII\,, and \FeII\ specifically indicates a localized contribution from a faster jet. The HV component, particularly in \NeIII\ is significant, as neon ionization requires energies $>$40\,eV, typically achieved only via fast dissociative $J$-type shocks ($>$80\,\kms) \citep{hollen89, shang10}, as strong X-ray/UV photoionization is unlikely in L1527. The inferred jet velocity ($>$100\,\kms) in L1527 is consistent with predictions of a shock-excited jet emerging from the inner disk, suggesting the existence of a HV jet core. Together, these observations support a stratified jet-outflow structure in which the fast jet core produces HV component in the ionized lines, while a moderate-velocity $J$-type shock with intermediate UV field contributes to the broader LV component throughout the outflow.

\begin{figure}[h]
\begin{center}
\includegraphics[width=1.03\linewidth]{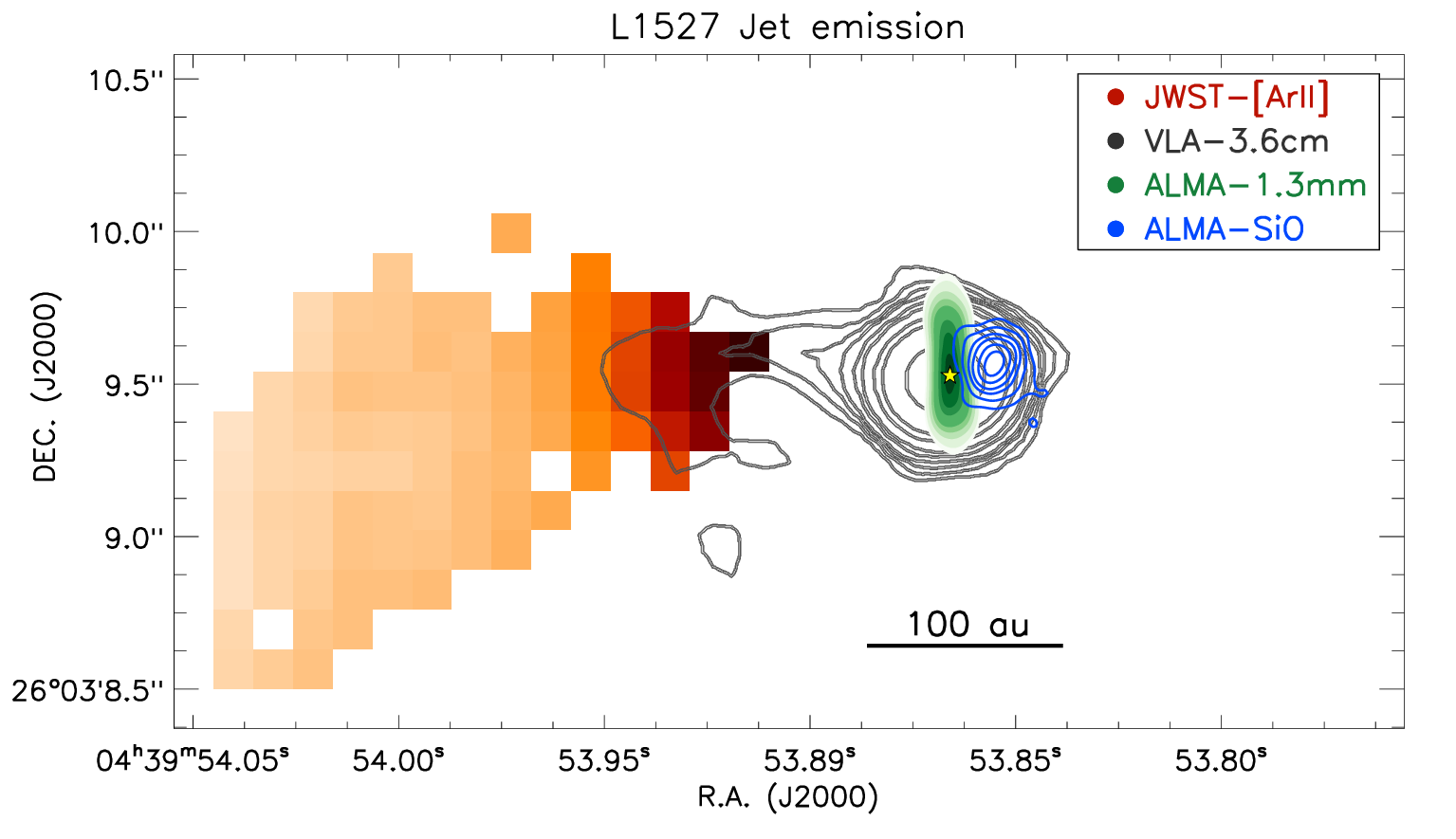}
 \caption{Complete map of jet emission in L1527. The red color shows the JWST \ArII\ HV component jet emission. The black contours show the VLA 3.6\,cm thermal emission \citep{reipurth04}. The green color shows the ALMA 1.3\,mm disk emission. The blue contours show the ALMA SiO jet emission \citep{vanthoff23}. The position of the central star is shown with a star.}
\label{fig11}
\end{center}
\end{figure}

Radio observations have been used to detect jets based on the continuum emission arising from the thermal Bremsstrahlung radiation (free-free emission) \citep[e.g.,][]{anglada98,furuya03}. An earlier study of L1527 by \citet{reipurth04} using the VLA "A" configuration at 3.6\,cm observed an extended continuum jet in the eastern region close to the protostar. The jet emission was extended to about 150\,au and was seen to curve toward the south at its leading edge (see the black contours in Fig.~\ref{fig11}). This detection was not taken into consideration, because it was suggested that it may be due to a binary, as proposed by \citet{loinard02}. However, based on recent high-sensitivity observations, it was shown that L1527 is not a binary \citep[e.g.,][]{nakatani20,sheehan22}. Correlating the morphology of the radio jet to the MIRI jet, we find similarity in the distribution. The jet axis matches spatially, extending toward the east and bending or curving toward the south. The curving of the jet could be due to either wiggling and precession, or an interaction with a dense layer of gas. A key feature that can be interpreted from both observations is the proper motion of the jet. The radio jet extends approximately to about 150\,au, whereas the MIRI jet emission extends up to 500\,au. There is a difference of about 350\,au in length between both the emissions, which have a time difference of 20 years. Based on this, a rough estimate of the average tangential velocity of the jet would lead to around $83\,$\kms. A main caveat in this estimate is that both datasets are observed at different wavelengths; therefore, the difference in the jet length may also be associated with the emission from different layers of gas.  

More recently, \citet{vanthoff23} carried out ALMA observations and detected a noticeable SiO emission in the western side (see the blue contours in Fig.~\ref{fig11}). SiO predominantly traces high-density ejection material associated with jets in the innermost regions of class 0/I protostars \citep[e.g., HH\,212;][]{lee20}. The SiO emission observed in L1527 is compact and positioned very close to the protostar ($\sim$10-50\,au). Most importantly, it is blueshifted and exhibits higher velocity when compared to ALMA CO observations. When considered alongside our MIRI jet emission in the eastern side, this detection strongly supports the bipolar nature of the jet in L1527. In addition, their velocities and positions confirm that the eastern region is redshifted while the western region is blueshifted consistent with the \NeII\ velocity maps (see Fig.~\ref{fig6}). \citet{vanthoff23} also examined a large-scale $^{12}$CO map and found hints of a collimated CO emission in the western side at a distance of around 500-800\,au. This emission appeared more resolved, flowing along the outflow axis, but had lower velocity than the SiO emission. Therefore, it could not be confirmed as a large-scale jet.

While the L1527 outflow exhibits a clear bipolar structure, the jet emission traced by the HV component and SiO appears asymmetric and poorly distributed. The confinement of the HV component to one region may reflect the effects due to combination of inclination, asymmetries in the medium, and/or an episodic accretion. Asymmetric jets have been observed in other deeply embedded sources, such as L1448-C \citep{dionatos09}, BHR\,71 \citep{gusdorf15}, HH\,46/47 \citep{zhang16}, and TMC\,1 \citep{tycho24}, where external density gradients or variable accretion lead to predominantly one-sided jet activity. The inhomogeneity of the medium significantly changes the excitation conditions leading to emission in only certain tracers for each region. Many theoretical simulations lend strong support to this interpretation. For instance, \citet{raga02} modeled jets in young stars and showed how early-stage jets can produce compact, asymmetric jet features, particularly when interacting with clumpy or inhomogeneous envelopes. \citet{hansen16} conducted 3D MHD simulations of class 0 jets and found that early jet interactions with over-dense clumps can produce complex asymmetric shock structures. In addition, such interactions also produce a number of reverse shocks along the flow, which seems to be the case in L1527 (see Fig.~\ref{fig1}). Moreover, it is believed that magnetic fields play a crucial role in shaping the jet structures \citep{cerqueira01,pudritz19}. Variations in magnetic field topology can lead to jets with uneven or asymmetric outflows. Different configurations of magnetic fields, such as misalignments between the field and the rotation axis due to a warped disk, or nonuniform field structures, can cause one side of the jet to be less collimated or more powerful than the other.

Another factor attributing to L1527's jet morphology is its evolutionary stage. While L1527 is classified as a class 0/I transitional object, its jet appears unusually weak and compact. This indicates that jet activity has either declined or undergoes episodes of periodic mass loss. Such temporal evolution has been seen in sources such as L1448-C, IRAS\,16293-2422, and Serpens South C7, where compact jets coexist with broader molecular flows and signs of episodic accretion are common \citep{bachiller90,kristensen13,plunkett15}. Spitzer IRAC observations of L1527 taken 12 months apart show total flux variability and emission variability in each outflow lobe \citep{tobin08}. Such variations are typically associated with asymmetric changes in accretion. The mass accretion rates inferred from modeling L1527 show it to be $\macc \approx 3-7\times10^{-7}\,\msun\mathrm{yr^{-1}}$ \citep{tobin10,flores21}. These values are two orders of magnitude higher when compared to values from class II protostars, indicating that L1527 is still young. Recent JWST NIRspec observational estimates on mass accretion rate show it to be $\macc \approx1.0\times10^{-7}\,\msun\mathrm{yr^{-1}}$ and support a nonsteady and asymmetric accretion scenario in L1527 \citep{drechsler26}.
 
A key morphological characteristic of L1527, compared to other class 0/I protostars, is the poor collimation of its outflow and jet emission. It is still unclear whether jet collimation is a factor of age or environment. However, the common consensus is that as protostars evolve over time, the degree of collimation tends to decrease. This trend is also accompanied by changes in jet properties, such as the transition from predominantly molecular to primarily ionized emission \citep{frank14, bally16,mottram17}. In the case of L1527, we do not see any evidence of a molecular jet in the MIRI/MRS H$_{2}$-integrated intensity maps (see Fig.~\ref{fig3}). Instead, most of the molecular emission traces the poorly collimated disk wind. We measured the outflow opening angle using the MIRI broadband images by tracing the cavity walls from about 2000\,au down to less than than 50\,au near the protostar. The estimated opening angle is about $84^{\circ}\pm3$, with no inclination correction as the outflow lies nearly in the plane of sky. This is consistent with the previous estimate of $85^{\circ}$ by \citet{tobin08} using Spitzer IRAC images. \citet{dunham24} studied the opening angles of 46 outflows in the Perseus cloud and found that outflows widen during the class 0 phase as they clear dense envelope material, but reach a $90^{\circ}$ limit by class I when most of the envelope has been dispersed and further expansion is no longer supported.

The unusually wide opening angle and lack of a well-collimated molecular jet make L1527 a unique jet-launching system and suggest either an atypical evolution pathway or a combination of MHD processes and environmental conditions. In particular, magnetocentrifugal winds are thought to self-collimate due to stresses generated by magnetic field lines anchored in the disk or the disk-star interface \citep{shu94, pudritz07}. A strong poloidal magnetic field near the disk is responsible for the initial launch and collimation. As the flow propagates, disk rotation twists the field lines, generating a toroidal magnetic field that further collimates the outflow at larger distances \citep{shang06}. When the poloidal field near the disk is weak due to turbulence, envelope infall, or disk misalignment, the initial outflow is poorly collimated, and the subsequent buildup of toroidal fields may not be sufficient to produce strong collimation. Numerical simulations \citep[e.g.,][]{tomisaka98,hennebelle08,joos12} demonstrate that when the poloidal field strength is low or the field strongly misaligns with the rotation axis, the resulting outflows are wide and poorly collimated.

\citet{cox15} carried out subarcsecond polarization observations of L1527 at 1.3\,mm and found a dominant toroidal field in the region without any poloidal component. In addition, they found the magnetic field to be misaligned with the rotation axis. However, recent high-sensitivity polarization studies suggest that such polarized emission may primarily trace scattered light rather than the intrinsic magnetic field structure \citep{harrison21,lin24}. Nevertheless, if taken as a direct tracer, this result combined with our observed morphology is consistent with a scenario of a weak misaligned poloidal field responsible for poor collimation. The strength of the magnetic field in the region also plays an important role in the degree of collimation. Several observations using dust polarization and Zeeman splitting have shown that class 0/I protostars with highly collimated jets have magnetic field strengths on the order of 1-100\,mG at scales less than 100\,au, namely IRAS\,16293-2422, HH\,211, and L1448 IRS2 \citep{alves12,lee18,kwon19}. In contrast, in protostars such as IRAS\,15398-3359 \citep{redaelli19}, where the outflows are less collimated, inferred field strengths at similar scales are lower, on the order of 10-100\,$\mu$G. So far there has not been any direct measurement of magnetic field strength in L1527. However, \citet{davidson14} tested different magnetic field models with observed polarization configurations to reproduce the observed disk and outflow morphology. They found that only the weakly aligned magnetic field model with strengths of about $100\,\mu$G favors the observed morphology. Therefore, it is likely that the magnetic field in L1527 is weak and contributes to poor collimation, leading to broad, wide-angle flows rather than narrow jets. These properties in L1527 suggest that even in the earliest embedded stages, variations in magnetic field configuration, in addition to environmental conditions, strongly influence the collimation and shaping of the outflow structures.

%Ne and Te from FeII or NeII lines.
%Correct the reference line of 140 au in all plots

\section{Summary and conclusions}
\label{summary}

We presented JWST MIRI/MRS spectral observations toward a young low-mass class 0/I protostar: L1527. This study has provided a detailed characterization of the physical and kinematic properties within the innermost 500\,au region of the L1527 outflow. Through high-resolution spectroscopic observations, we identified several molecular, atomic, and ionized lines tracing distinct outflow components. The most striking result is a previously undetected, short, and poorly collimated HV ionized jet in the eastern region, which indicates active accretion onto the protostar. Conversely, no molecular H$_{2}$ jet is detected at smaller scales in the MIRI data, nor a CO jet at larger scales in other studies, making the L1527 outflow a unique among outflows from young stars. Below, we summarize the main results of our study.\\
\indent - Broadband MIR imaging of L1527 reveals a bright bipolar structure consisting of several filaments, bow shocks, knots, and reverse shocks, dominated by scattered light. The outflow lobes are significantly bright and have a large opening angle of $84^{\circ}\pm3$.\\
\indent - The MIRI/MRS spectra reveal a chemically rich environment composed of molecular, atomic, and ionized species. Pure rotational H$_{2}$ $\nu =$(0-0) transitions from S(1) to S(8) are observed, with S(4) and S(5) being the strongest. The lines appear associated with a single component of the outflow, most likely the outflow cavity wall.\\
\indent - Forbidden lines of \NiII, \ArII, \NeII, \NeIII, \SI, and \FeII, were detected tracing distinct ionized layers within the outflow. \ArII, \NeII, and \FeII\ show widespread shock excitation.\\
\indent - \NeII\ channel maps highlight spatially distinct emission regions, indicating shock activity. The radial velocity map shows that the eastern lobe is redshifted and the western lobe is blueshifted, contrary to earlier interpretations. The average radial velocity is $\sim30$\,\kms.\\
\indent - Double-peaked emission profiles in the \ArII, \NeIII, and \FeII\ lines were detected and found confined only to the eastern region. The LV ($-$30\,\kms) component traces the material associated with the general disk wind, while the HV (100\,\kms) component traces a poorly collimated fast jet extending up to $\sim$500\,au from the protostar.\\
\indent - An excitation diagram analysis of H$_{2}$ rotational lines reveals two distinct gas layers: a warm component ($T_\mathrm{warm}\approx$550\,K, $N(\rm{H_{2}})_\mathrm{warm}\approx 10^{20}$\,cm$^{-2}$) and a hot component ($T_\mathrm{hot}\approx$2500\,K, $N(\rm{H_{2}})_\mathrm{hot}\approx 10^{18}$\,cm$^{-2}$). The spatial variations in temperature and column density maps closely follow the outflow morphology in H$_{2}$ emission, suggesting that gas heating and compression are strongly linked to a common excitation mechanism. \\
\indent - A comparison of molecular and ionized emission with shock models indicates that moderate-velocity $J$-shocks with a modest UV field are the dominant drivers of the widespread excitation seen in the observed lines.\\

Overall, the JWST MIRI/MRS observations of L1527 offer new insights into the early phases of protostellar evolution. The data reveal a chemically rich environment shaped by dynamic interactions and shocks. The outflow exhibits a stratified structure of molecular and ionized gas, where a compact jet coexists with a slower wide-angle disk wind, likely driven by weak magnetic fields. These findings suggest that stratified wide-angle outflows are active even in deeply embedded stages, challenging assumptions that such features are confined to later stages.
Future high-spatial-resolution observations combined with magnetic field studies will be essential to further constrain the origin and dynamics of the outflow and jet structures.
 
\begin{acknowledgements}

This work is based on observations made with the NASA/ESA/CSA James Webb Space Telescope. The data were obtained from the Mikulski Archive for Space Telescopes at the Space Telescope Science Institute, which is operated by the Association of Universities for Research in Astronomy, Inc., under NASA contract NAS 5-03127 for JWST. These observations are associated with program PID 1290 and PID 1798.
The data described here may be obtained from \url{https://doi.org/10.17909/5dw8-cs47}.

This research has been supported by the European Research Council advanced grant H2020-ER-2016-ADG-743029 under the European Union’s Horizon 2020 Research and Innovation program. 

This paper makes use of the following ALMA data: ADS/JAO.ALMA 2019.1.00261.L and ADS/JAO.ALMA 2019.A.00034.S. ALMA is a partnership of ESO (representing its member states), NSF (USA) and NINS (Japan), together with NRC (Canada), MOST and ASIAA (Taiwan), and KASI (Republic of Korea), in cooperation with the Republic of Chile. The Joint ALMA Observatory is operated by ESO, AUI/NRAO and NAOJ. 

EvD, LT, LF and MvG acknowledge support from ERC Advanced grant 101019751 MOLDISK, TOP-1 grant 614.001.751 from the Dutch Research Council (NWO), the Netherlands Research School for Astronomy (NOVA), the Danish National Research Foundation through the Center of Excellence "InterCat" (DNRF150), and DFGgrant 325594231, FOR 2634/2.

A.C.G. acknowledges support from PRIN-MUR 2022 20228JPA3A "The path to star and planet formation in the JWST era (PATH)" funded by NextGeneration EU and by INAF-GoG 2022 "NIR-dark Accretion Outbursts in Massive Young stellar objects (NAOMY)" and Large Grant INAF 2022 "YSOs Outflows, Disks and Accretion: towards a global framework for the evolution of planet forming systems (YODA)". 

H.B. acknowledges support from the Deutsche Forschungsgemeinschaft in the Collaborative Research Center (SFB 881) "The Milky Way System" (subproject B1).

P.J.K. acknowledges financial support from the Science Foundation Ireland/Irish Research Council Pathway program under Grant Number 21/PATH-S/9360.

The following National and International Funding Agencies funded and supported the MIRI development: NASA; ESA; Belgian Science Policy Office (BELSPO); Centre Nationale d’Etudes Spatiales (CNES); Danish National Space Centre; Deutsches Zentrum fur Luft- und Raumfahrt (DLR); Enterprise Ireland; Ministerio De Economia y Competividad; Netherlands Research School for Astronomy (NOVA); Netherlands Organization for Scientific Research (NWO); Science and Technology Facilities Council; Swiss Space Office; Swedish National Space Agency; and UK Space Agency.

The authors acknowledge the use of SAOImage DS9 software which is developed with the funding from the Chandra X-ray Science Center, the High Energy Astrophysics Science Archive Center and JWST Mission office at Space Telescope Science Institute. 
     
\end{acknowledgements}

\begin{appendix}

\onecolumn
\section{MIRI/MRS spectral lines in the L1527 outflow}
Table~\ref{tabA1} and \ref{tabA2} summarizes the list of detected molecular, atomic and ionized lines in the L1527 outflow with line intensities obtained at aperture positions in the eastern and western lobes.

\begin{table*}[h!]
\caption{Properties of the list of detected molecular H$_2$ pure rotational lines. }
\label{tabA1}
\centering
\begin{tabular}{cccclcc}
\hline \hline
Molecular &  Wavelength & Excitation Energy & Statistical & MIRI Channel- & \multicolumn{2}{c}{Line Intensity (W\,m$^{-2}$\,arcsec$^{-2}$)} \\
Line & $\lambda$ ($\mu$m) & $E_\mathrm{u}$/$k_\mathrm{B}$ (K) & Weight $g_{u}$ & Grating  & EAST & WEST \\
\hline
H$_2$\,(0-0) S(8) & 5.05312 & 8677  & 21 & ch1-SHORT & $1.57\times10^{-20}$  & $2.11\times10^{-20}$ \\
H$_2$\,(0-0) S(7) & 5.51116 & 7197  & 57 & ch1-MEDIUM & $5.08\times10^{-19}$  & $4.07\times10^{-19}$ \\
H$_2$\,(0-0) S(6) & 6.10856 & 5830  & 17 & ch1-MEDIUM & $5.45\times10^{-20}$  & $4.87\times10^{-20}$ \\
H$_2$\,(0-0) S(5) & 6.90952 & 4586  & 45 & ch1-LONG & $7.61\times10^{-19}$  & $6.57\times10^{-19}$ \\
H$_2$\,(0-0) S(4) & 8.02505 & 3474  & 13 & ch2-SHORT & $7.08\times10^{-19}$  & $6.82\times10^{-19}$ \\
H$_2$\,(0-0) S(3) & 9.66491 & 2504  & 33 & ch2-MEDIUM & $5.39\times10^{-20}$  & $4.23\times10^{-20}$ \\
H$_2$\,(0-0) S(2) & 12.2786 & 1682  & 9 & ch3-SHORT & $4.01\times10^{-20}$  & $4.72\times10^{-20}$ \\
H$_2$\,(0-0) S(1) & 17.0348 & 1015  & 21 & ch3-LONG & $7.83\times10^{-20}$  & $8.05\times10^{-20}$ \\
 \hline
\end{tabular}
\end{table*}

\begin{table*}[ht]
\begin{center}

\caption{Properties of the list of detected atomic and ionized lines.}
\label{tabA2}
\begin{tabular}{ccccclcc}
\hline
\hline
Ion & Wavelength &  Excitation Energy & Ionization & Line ID  & MIRI Channel- & \multicolumn{2}{c}{Line Intensity (W\,m$^{-2}$\,arcsec$^{-2}$)} \\ 
& $\lambda$ ($\mu$m)  & $E_\mathrm{u}$/$k_\mathrm{B}$ (K) & Potential (eV) &   &  Grating & EAST & WEST \\
\hline
\FeII  & 5.341  & 2694  & 7.90   &  a$^4$F$_{9/2}$-a$^6$D$_{9/2}$ &  ch1-SHORT & $3.84\times10^{-19}$  & $0.94\times10^{-19}$ \\
\NiII  & 6.636  & 2168  & 7.64   &  $^2$D$_{3/2}$-$^2$D$_{5/2}$   &  ch1-LONG  &   					 $-$  & $2.43\times10^{-20}$ \\
\ArII  & 6.985  & 2060  & 15.76  &  $^2$P$_{1/2}$-$^2$P$_{3/2}$   &  ch1-LONG  & $8.94\times10^{-19}$  & $-$ \\ %fund
\NeII  & 12.813 & 1123  & 21.56  &  $^2$P$_{1/2}$-$^2$P$_{3/2}$   &  ch3-SHORT & $3.17\times10^{-18}$  & $2.51\times10^{-19}$ \\ %fund
\NeIII & 15.555 & 925   & 40.96  &  $^3$P$_{2}$-$^3$P$_{1}$       &  ch3-SHORT & $3.75\times10^{-19}$  & $1.13\times10^{-19}$ \\ %fund
\FeII  & 17.936 & 3497  & 7.90   &  a$^4$F$_{7/2}$-a$^4$F$_{9/2}$ &  ch3-LONG & $2.06\times10^{-19}$   & $9.86\times10^{-20}$ \\
\SI    & 25.249 & 570   & 0      &  $^3$P$_{1}$-$^3$P$_{2}$       &  ch4-LONG & $4.51\times10^{-20}$   & $6.28\times10^{-19}$ \\
\FeII  & 25.988 & 554   & 7.90   &  a$^6$D$_{7/2}$-a$^6$D$_{9/2}$ &  ch4-LONG & $1.93\times10^{-20}$   & $2.25\times10^{-20}$ \\
 \hline
\end{tabular}
\tablefoot{Blank entries in the eastern and western line intensity columns indicate non-detection or intensity levels below the noise threshold.}

%$^b$ Excitation temperature of the upper level\\
%$^a$ Ionization potential of the $X^{i-1}$ ion \\
\end{center}
\end{table*}

\newpage

\begin{figure}
\section{Excitation diagram analysis}

Figures~\ref{figb1} and \ref{figb3} show the excitation diagram analysis results with the extinction-corrected H$_{2}$ line intensities and two-component fits for observed values at aperture positions in the eastern and western lobes of L1527 outflow.

\begin{center}
\includegraphics[width=1.0\linewidth]{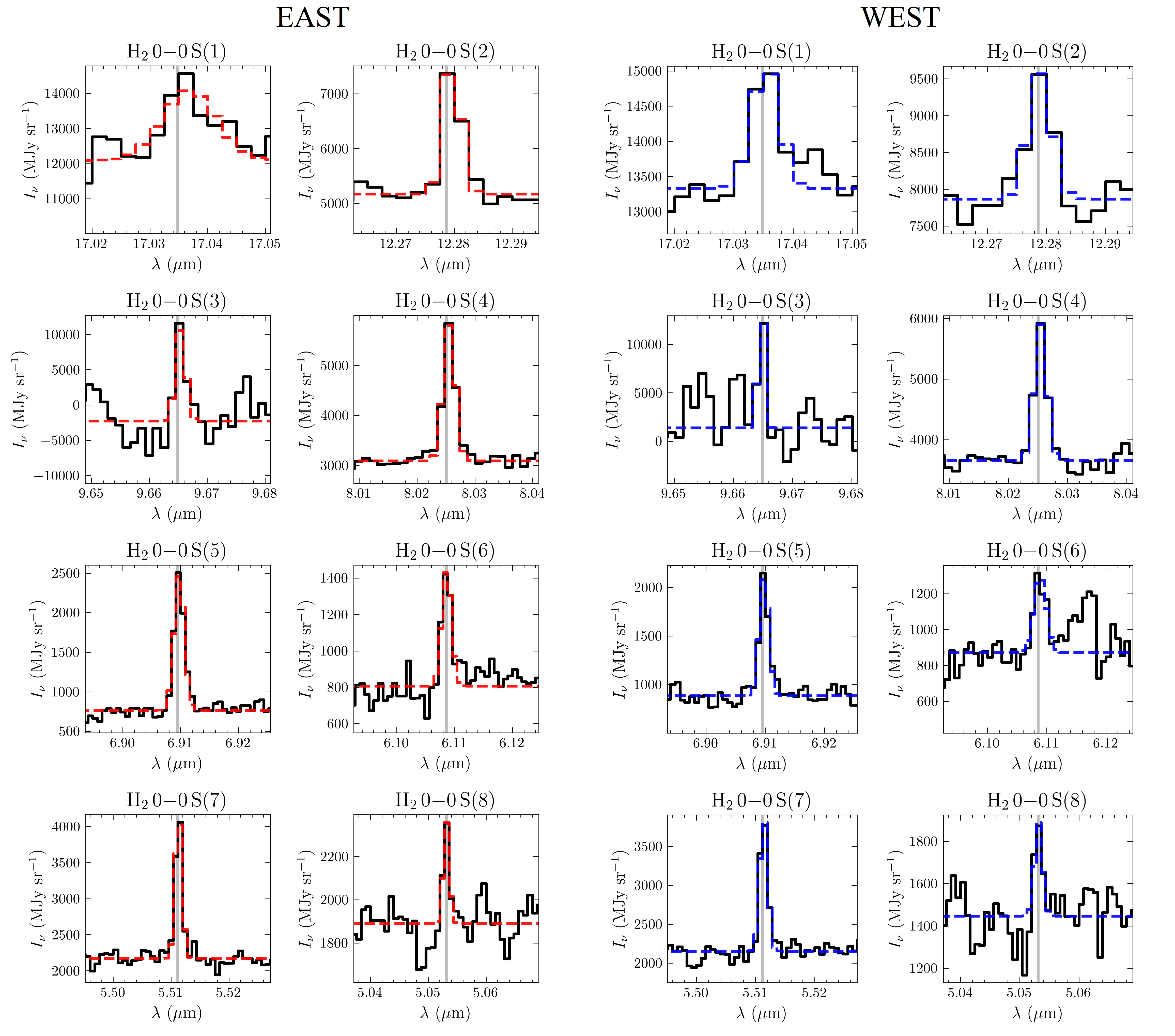}
\caption{Extinction-corrected H$_{2}$ rotational lines from S(1) to S(8) transitions for example spaxels at aperture positions in the eastern and western lobes (see Fig.~\ref{fig1}). The Gaussian fits to the lines are color-coded in red and blue, respectively, for each region. The vertical gray line indicates the central wavelength of each transition.}
  \label{figb1}
  \end{center}
\end{figure}

\begin{figure}[h]
\begin{center}
\includegraphics[width=0.5\linewidth]{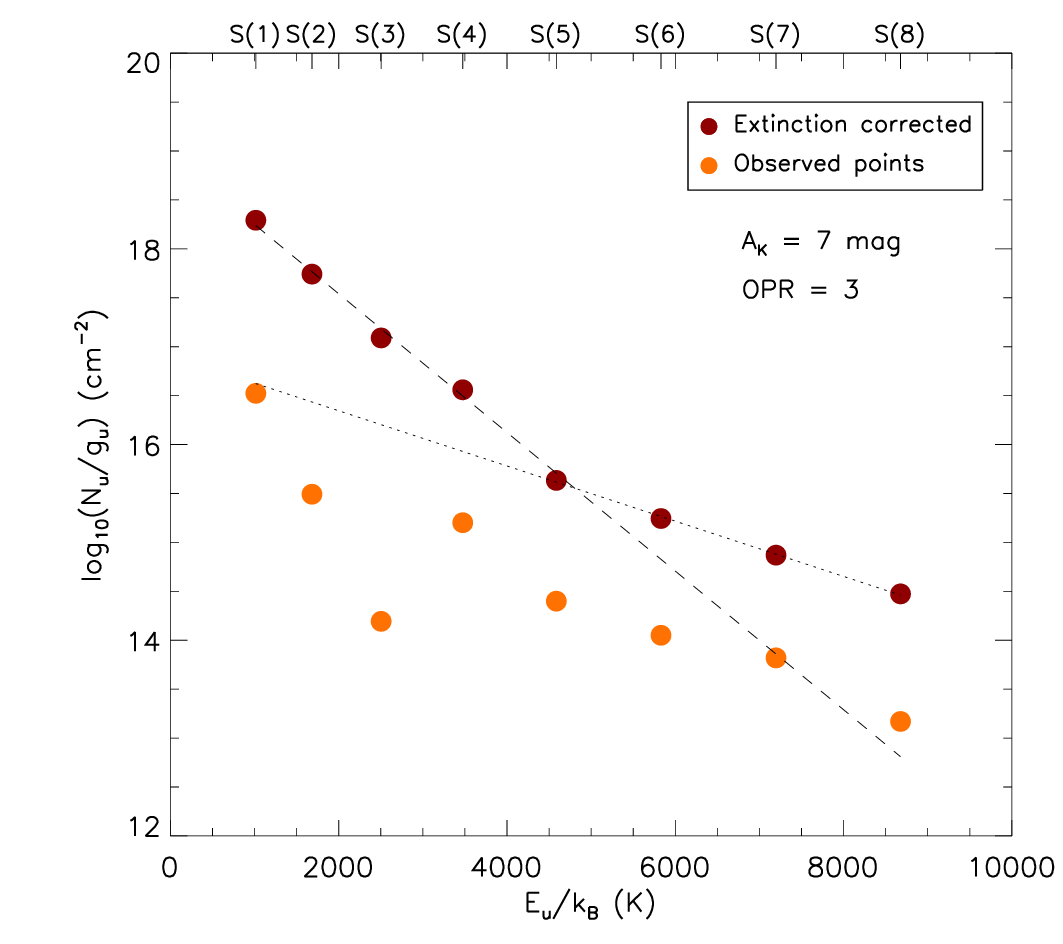}
 \caption{H$_{2}$ excitation diagram before and after extinction correction for transitions from S(1) to S(8) obtained at aperture position in the eastern lobe. The extinction correction of A$_{K}$=7\,mag best represents the two-component fit.}
\label{figb2}
\end{center}
\end{figure}

\begin{figure}
\begin{center}
\includegraphics[width=1.0\linewidth]{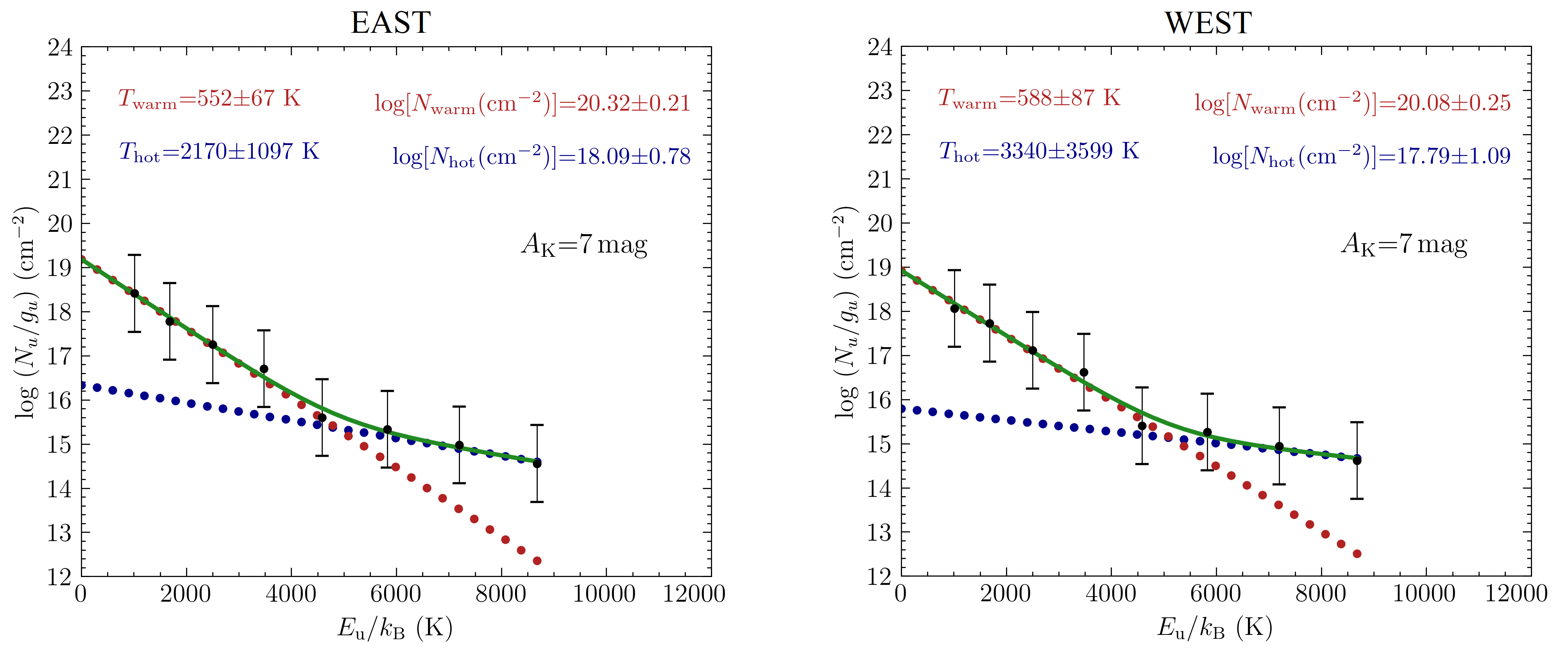}
\caption{H$_{2}$ excitation diagram analysis using \texttt{pdrtpy} for example spaxels at aperture positions in the eastern and western lobes (see Fig.~\ref{fig1}). Extinction-corrected line intensities, converted to upper-state column densities, are shown as black data points with error bars. The two-component fit to the data points is shown by red and blue dots, for the warm and hot component, respectively. The total fit is indicated by a green line. Estimated temperature and column density values are shown at the top. The errors bars are estimated from uncertainties in the line-integrated intensities, absolute flux calibration (5\%), and extinction (2\,mag). These uncertainties are then propagated to the upper-state column densities.}
  \label{figb3}
\end{center}
\end{figure}

\end{appendix}

\begin{thebibliography}{}
%
\footnotesize

\bibitem[Agra-Amboage et al.(2011)]{agra11}
Agra-Amboage, V., Dougados, C., Cabrit, S. and Reunanen, J., 2011, \href{http://doi.org/10.1051/0004-6361/201015886} \aap, 532, A59

\bibitem[Alves et al.(2012)]{alves12}
Alves, F. O., Vlemmings, W. T., Girart, J. M., \& Torrelles, J. M. 2012, \href{https://doi.org/10.1051/0004-6361/201118710} \aap, 542, A14

\bibitem[Anglada et al.(1998)]{anglada98}
Anglada, G., Villuendas, E., Estalella, R., M. T., Beltran, M., et al. 1998, \href{https://doi.org/10.1086/300637} \aj, 116, 2953

\bibitem[Anglada et al.(2018)]{anglada18}
Anglada G., Rodriguez, L. F., Carrasco-Gonzalez, C., 2018, \href{https://doi.org/10.1007/s00159-018-0107-z} A\&ARv, 26, 3

\bibitem[Arce et al.(2013)]{arce13}
Arce, H. G., Mardones, D., Corder, S., Garay, G., et al. 2013, \href{http://doi.org/10.1088/0004-637X/774/1/39} \apj, 774, 1

\bibitem[Argyriou et al.(2023)]{argyriou23}
Argyriou, I., Glasse, A., Law, D. R., Labiano, A., et al. 2023, \href{https://doi.org/10.1051/0004-6361/202346489} \aap, 675, A111

\bibitem[Aso et al.(2013)]{aso17}
Aso, Y., Ohashi, N., Aikawa, Y., et al. 2017, \href{https://doi.org/10.3847/1538-4357/aa8264} \apj, 849, 56

\bibitem[Bachiller et al.(1990)]{bachiller90}
Bachiller, R., Cernicharo, J., Martin-Pintado, J., Tafalla, M., \& Lazareff, B. 1990, \href{http://linker.aanda.org/10.1051/0004-6361/201937046/7} \aap, 231, 174

\bibitem[Banerjee \& Pudritz (2006)]{ban06}
Banerjee, R., Pudritz, R. E., 2006, \href{https://doi.org/10.1086/500496} \apj, 641, 949

\bibitem[Banerjee \& Pudritz (2007)]{ban07}
Banerjee, R., Pudritz, R. E., 2007, \href{https://doi.org/10.1086/512010} \apj, 660, 479

\bibitem[Bally (2016)]{bally16}
Bally, J., 2016, \href{https://doi.org/10.1146/annurev-astro-081915-023341} \araa, 54, 491

\bibitem[Beuther et al.(2023)]{beuther23}
Beuther, H., van Dishoeck, E., Tychoniec, L., Gieser, C., Kavanagh, P. J., et al. 2023, \href{https://doi.org/10.1051/0004-6361/202346167} \aap, 673, A121

\bibitem[Blandford \& Payne (1982)]{blan82}
Blandford, R. D., Payne, D. G., 1982, \href{https://doi.org/10.1093/mnras/199.4.883} \mnras, 199, 883

\bibitem[Bitner (2008)]{bitner08}
Bitner, M. A., Richter, M. J., Lacy, J. H., et al. 2008, \href{https://doi.org/10.1086/592317} \apj, 688, 1326

\bibitem[Beverly (1962)]{lynds62}
Beverly, Lynds, 1962, \href{https://doi.org/10.1086/190072} \apjs, 7, 1L

\bibitem[Bushouse et al.(2023)]{bushouse23}
Bushouse, H., Eisenhamer, J., Nadia, D., Davies, J., et al. 2023,\href{https://zenodo.org/doi/10.5281/zenodo.10022973} Zenodo, JWST calibration pipeline.

\bibitem[Bontemps et al.(1996)]{bontemps96}
Bontemps, S., Andre, P., Terebey, S., Cabrit, S., 1996, \href{http://linker.aanda.org/10.1051/0004-6361/202451291/16} \aap, 311, 858

\bibitem[Caratti o Garatti et al.(2024)]{caratti24}
Caratti o Garatti, A., Ray, T. P., Kavanagh, P., McCaughrean, M. J., et al. 2024, \href{https://doi.org/10.1051/0004-6361/202451350} \aap, 691, A134

\bibitem[Cerqueira \& Gouveia Dal Pino (2001)]{cerqueira01}
Cerqueira, A. H. \& Gouveia Dal Pino, E. M., 2001, \href{https://doi.org/10.1086/322245} \apj, 560, 779

\bibitem[Christiaens et al.(2023)]{christiaens23}
Christiaens, V., Gonzalez, C., Farkas, R., et al. 2023, \href{http://linker.aanda.org/10.1051/0004-6361/202348118/22} J. Open Source Software, 8, 4774

\bibitem[Crouzet et al.(2025)]{crouzet25}
Crouzet, N., Mueller, M., Sargent, B., Lahuis, F., Kester, D., et al. 2025, \href{	https://doi.org/10.1051/0004-6361/202452903} \aap, 698, A77

\bibitem[Davidson et al.(2014)]{davidson14} 
Davidson, J., Li, Z.-Y., Hull, C., Plambeck, R., Kwon, W., Crutcher, R., et al. 2011, \href{http://doi.org/10.1088/0004-637X/797/2/74} \apj, 979, 74

\bibitem[Davis et al.(2010)]{davis10} 
Davis, C. J., Gell, R., Khanzadyan, T., Smith, M. D., Jenness, T. 2010, \href{http://doi.org/10.1051/0004-6361/200913561} \aap, 511, A24

\bibitem[Davis et al.(2011)]{davis11} 
Davis, C. J., Cervantes, B., Nisini, B., Giannini, T., et al. 2011, \href{https://doi.org/10.1051/0004-6361/201015897} \aap, 528, A3

\bibitem[Devaraj et al.(2023)]{dev23} 
Devaraj, R., Caratti o Garatti, A., Ray, T. P., Dewangan, L. K., et al. 2023, \href{https://doi.org/10.3847/1538-4357/acb68e} \apj, 944, 226

\bibitem[Delabrosse et al.(2024)]{delab24} 
Delabrosse, V., ,Dougados, C., Cabrit, S., Tabone, B., et al. 2024, \href{https://doi.org/10.1051/0004-6361/202449176} \aap, 688, A173

\bibitem[Dionatos et al.(2009)]{dionatos09} 
Dionatos, O., Nisini, B., Garcia Lopez, R., Giannini, T., Davis, C. J., et al. 2009, \href{http://doi.org/10.1088/0004-637X/692/1/1} \apj, 692, 1

\bibitem[Drechsler et al.(2026)]{drechsler26} 
Drechsler, B., Tobin, J., Sheehan, P., Looney, L., et al. 2026, \apj, Submitted

\bibitem[Dunham et al.(2014)]{dunham14} 
Dunham, M. M., Arce, H. G., Mardones, D., et al. 2014, \href{http://doi.org/10.1088/0004-637X/783/1/29} \apj, 783, 29

\bibitem[Dunham et al.(2024)]{dunham24} 
Dunham, M., Stephens, I., Myers, P., Bourke, T., Arce, H. G., et al. 2024, \href{https://doi.org/10.1093/mnras/stae2018} \mnras, 533, 4

\bibitem[Eisl\"{o}ffel et al.(2000)]{eisloffel00}
Eisl\"{o}ffel, J., Mundt, R., Ray, T. P., \& Rodriguez, L. F. 2000, \href{https://ui.adsabs.harvard.edu/abs/2000prpl.conf..815E/abstract} PPIV, Univ. of Arizona Press), 815

\bibitem[Frank et al.(2014)]{frank14}
Frank, A., Ray, T. P., Cabrit, S., Hartigan, P., Arce, H. G., et al. 2014, \href{https://ui.adsabs.harvard.edu/link_gateway/2014prpl.conf..451F/doi:10.2458/azu_uapress_9780816531240-ch020} PPVI, Univ. of Arizona Pres, 451 

\bibitem[Federrath (2015)]{federrath15}
Federrath, C., 2015, \href{https://doi.org/10.1093/mnras/stv941} \mnras, 450, 4035

\bibitem[Ferreira (1997)]{ferreira97}
Ferreira, J., 1997, \href{https://ui.adsabs.harvard.edu/link_gateway/1997A&A...319..340F/doi:10.48550/arXiv.astro-ph/9607057} \aap, 319, 340

\bibitem[Flores-Rivera et al.(2021)]{flores21}
Flores-Rivera, L., Terebey, S., Willacy, K., Isella, A., et al. 2021, \href{https://doi.org/10.3847/1538-4357/abd1db} \apj, 908, 108

\bibitem[Flower \& Pineau des Fort\^{e}s (2003)]{flower03}
Flower, D. R. \& Pineau des For\^{e}ts, G., 2003, \href{https://doi.org/10.1046/j.1365-8711.2003.06716.x} \mnras, 343, 390

\bibitem[Flower \& Pineau des For\^ets (2015)]{flower15}
Flower, D. R., \& Pineau des For\^ets, G. 2015, \href{https://doi.org/10.1051/0004-6361/201525740} \aap, 578, A63

\bibitem[Feeney-Johansson et al.(2019)]{anton19}
Feeney-Johansson, A., Purser, S. J. D., Ray, T. P., et al. 2019, \apj, 885, L7

\bibitem[Francis et al.(2025)]{francis25}
Francis, L., van Dishoeck, E., Caratti o Garatti, A., van Gelder, M. L., Gieser, C., et al. 2025, \href{https://doi.org/10.1051/0004-6361/202451629} \aap, 694, A174

\bibitem[Froebrich et al.(2003)]{froebrich03}
Froebrich, D., Smith, M., \& Eisloffel, J., 2003, \href{https://ui.adsabs.harvard.edu/link_gateway/2003Ap&SS.287..217F/doi:10.1023/B:ASTR.0000006227.85806.85} Astrophysics and Space Science, 287, p217

\bibitem[Furuya et al.(2003)]{furuya03}
Furuya R. S., Kitamura Y., Wootten A., Claussen M. J.,Kawabe R., 2003, \href{http://doi.org/10.1086/342749} \apjs, 144, 71

\bibitem[Galli et al.(2019)]{galli19}
Galli, P., Loinard, L.,Bouy, H., Sarro, L., et al. 2019, \href{https://doi.org/10.1051/0004-6361/201935928} \aap, 630, A137

\bibitem[Garcia et al.(2001)]{garcia01}
Garcia, P. V., Cabrit, S., Ferreira, J., Binette, L., 2001, \href{https://doi.org/10.1051/0004-6361:20011146} \aap, 377, 609

\bibitem[Garcia Lopez et al.(2008)]{garcia08}
Garcia Lopez, R., Nisini, B., Giannini, T., Eisloffel, J., Bacciotti, F., et al. 2008, \href{http://doi.org/10.1051/0004-6361:20079045} \aap, 487, 1019 

\bibitem[Garcia Lopez et al.(2010)]{garcia10}
Garcia Lopez, R., Nisini, B., Eisloffel, J., Giannini, T., Bacciotti, F., et al. 2010, \href{http://doi.org/10.1051/0004-6361/200913304} \aap, 511, A5 

\bibitem[Gardner et al.(2023)]{gardner23}
Gardner, J. P., Mather, J. C., Abbott, R., Abell, J. S., et al. 2023, \href{https://doi.org/10.1088/1538-3873/acd1b5} \pasp, 135, 068001

\bibitem[Giannini et al.(2001)]{giannini01}
Giannini, T., Nisini, B., Lorenzetti, D., 2001, \href{https://doi.org/10.1086/321451} \apj, 555, 40

\bibitem[Giannini et al.(2002)]{giannini02}
Giannini, T., Nisini, B., Caratti o Garatti, A., Lorenzetti, D., 2002, \href{https://doi.org/10.1086/340883} \apj, 570, L33

\bibitem[Gieser et al.(2023)]{gieser23}
Gieser, C., Beuther, H., van Dishoeck, E., Francis, L., van Gelder, M. L., et al. 2023, \href{https://doi.org/10.1051/0004-6361/202347060} \aap, 679, A108

\bibitem[Godard et al.(2019)]{godard19}
Godard, B., Pineau des For\^ets, G., Lesaffre, P., et al. 2019, \aap, 622, A100

\bibitem[Gomez-Ruiz et al.(2012)]{gomez12}
Gomez-Ruiz, A., Gusdorf, A., Leurini, S., Codella, C., et al. 2012, \href{http://doi.org/10.1051/0004-6361/201218936} \aap, 542, L9

\bibitem[G\"{u}del et al.(2010)]{gudel10}
G\"{u}del, M., Lahuis, F., Briggs, K. R., et al. 2010, \href{https://doi.org/10.1051/0004-6361/200913971} \aap, 519, A113

\bibitem[Gusdorf et al.(2015)]{gusdorf15}
Gusdorf, A., Riquelme, D., Anderl, S., Eisl\"{o}ffel, J., Codella, C., et al. 2015, \href{http://doi.org/10.1051/0004-6361/201425142} \aap, 575, A98

\bibitem[Guszejnov et al.(2022)]{guszejnov22}
Guszejnov, D., Grudic, M. Y., Offner, S. S. R., et al. 2022, \href{https://doi.org/10.1093/mnras/stac2060} \mnras, 515, 4929

\bibitem[Hansen et al.(2016)]{hansen16}
Hansen, E. C., Frank, A., Hartigan, P. \& Lebedev, S. V., 2016, \href{https://doi.org/10.3847/1538-4357/aa5ca8} \apj, 837, 143

\bibitem[Harrison et al.(2021)]{harrison21}
Harrison, R., Looney, L., Stephens, I., Li, Z-Y., Teague, R., et al. 2021, \href{https://doi.org/10.3847/1538-4357/abd94e} \apj, 908, 141

\bibitem[Hartigan et al.(1987)]{hartigan87}
Hartigan, P., Raymond, J., \& Hartmann, L. 1987, \href{https://doi.org/10.1086/165204} \apj, 316, 323

\bibitem[Hennebelle \& Fromang (2008)]{hennebelle08}
Hennebelle, P., \& Fromang, S., 2008, \href{https://doi.org/10.1051/0004-6361:20078309} \aap, 477, 9

\bibitem[Hsieh et al.(2023)]{hsieh23}
Hsieh, C-H., Arce, H., Li, Z-Y., Dunham, M., et al. 2023, \href{https://doi.org/10.3847/1538-4357/acba13} \apj, 947, 25

\bibitem[Hogerheijde et al.(1998)]{hoger98}
Hogerheijde M. R., van Dishoeck E. F., Blake G. A., van Langevelde H. J., 1998, \href{https://doi.org/10.1086/305885} \apj, 502, 315

\bibitem[Hollenbach \& McKee (1989)]{hollen89}
Hollenbach, D., \& McKee, C., 1989, \href{https://doi.org/10.1086/167595} \apj, 342, 306

\bibitem[Hollenbach \& Gorti (2009)]{hollen09}
Hollenbach, D., \& Gorti, U., 2009, \href{http://dx.doi.org/10.1088/0004-637X/703/2/1203} \apj, 703, 1203

\bibitem[Joos et al.(2012)]{joos12}
Joos, M., Hennebelle, P., \& Ciardi, A., 2012, \href{https://doi.org/10.1051/0004-6361/201118730} \aap, 543, A128

\bibitem[J\o rgensen et al.(2007)]{jorgensen07}
J\o rgensen, J., Bourke, T., Myers, P., Di Francesco, J., van Dishoeck, E., et al. 2007, \href{https://doi.org/10.1086/512230} \apj, 659, 479

\bibitem[Karska et al.(2018)]{karska18}
Karska, A., Kaufman, M. J., Kristensen, L. E., et al. 2018, \href{https://doi.org/10.3847/1538-4365/aaaec5} \apjs, 235, 30

\bibitem[Kristensen et al.(2011)]{kristensen13}
Kristensen, L. E., Klaassen, P., Mottram, J., Schmalzl, M., et al. 2013, \href{http://doi.org/10.1051/0004-6361/201220668} \aap, 549, L6

\bibitem[Kristensen et al.(2017)]{kristensen17}
Kristensen, L. E., van Dishoeck, E. F., Mottram, J. C., Karska, A., et al. 2017, \href{https://doi.org/10.1051/0004-6361/201630127} \aap, 605, A93

\bibitem[Kristensen et al.(2023)]{kristensen23}
Kristensen, L. E., Godard, B., Guillard, P., et al. 2023 \href{https://doi.org/10.1051/0004-6361/202346254} \aap, 675, A86

\bibitem[Kwon et al.(2019)]{kwon19}
Kwon, W., Stephens, I. W., Tobin, J. J., Looney, L. W., Li, Z.-Y., et al. 2019, \href{https://doi.org/10.3847/1538-4357/ab24c8} \apj, 879, 25

\bibitem[Law et al.(2023)]{law23}
Law, D. D., Morrison, J. E., Argyriou, I., Patapis, P., et al. 2023, \href{https://doi.org/10.3847/1538-3881/acdddc} \aj, 166, 45

\bibitem[Lee et al.(2002)]{lee02}
Lee, C.-F., Mundy, L. G., Stone, J., Ostriker, E., 2002, \href{https://doi.org/10.1086/341540} \apj, 576, 294

\bibitem[Lee et al.(2018)]{lee18}
Lee, C.-F., Hwang, H.-C., Ching, T.-C., Hirano, N., 2018, \href{https://doi.org/10.1038/s41467-018-07143-8} Nature, 9, 4636

\bibitem[Lee (2020)]{lee20}
Lee, C.-F. 2020, \href{https://doi.org/10.1007/s00159-020-0123-7} A\&A Rv, 28, 1

\bibitem[Le Gouellec (2025)]{gou25}
Le Gouellec, V., Lew, B., Greene, T., Johnstone, D., et al., 2025, \href{https://doi.org/10.3847/1538-4357/adcac4} \apj, 985, 225

\bibitem[Lin et al.(2024)]{lin24}
Lin, Z-Y. D., Li, Z-Y, Stephens, I., Fernandez-Lopez, M., et al. 2024, \href{https://doi.org/10.1093/mnras/stae040} \mnras, 528, 843

\bibitem[Liu et al.(2016)]{liu16}
Liu, C-F., Shang, H., Herczeg, G. J., and Walter F. M., 2016, \href{http://dx.doi.org/10.3847/0004-637X/832/2/153} \apj, 832, 153 

\bibitem[Loinard et al.(2002)]{loinard02}
Loinard, L., Rodriguez, L. F., D’Alessio, P., Wilner, D. J., \& Ho, P., 2002, \href{https://doi.org/10.1086/345940} \apj, 581, L109

\bibitem[Luhman (2018)]{luhman18}
Luhman, K., 2018 \href{https://doi.org/10.3847/1538-3881/aae831} \apj, 156, 271

\bibitem[Mottram et al.(2017)]{mottram17}
Mottram, J. C., van Dishoeck, E. F., Kristensen, L. E., Karska, A., et al. 2017, \href{https://doi.org/10.1051/0004-6361/201628682} \aap, 600, A99 

\bibitem[McClure (2009)]{mcclure09}
McClure, M. 2009, \href{http://doi.org/10.1088/0004-637X/693/2/L81} \apj, 693, L81

\bibitem[Nakatani et al.(2020)]{nakatani20}
Nakatani, R., Liu, H. B., Ohashi, S., et al. 2020, \href{https://doi.org/10.3847/2041-8213/ab8eaa} \apjl, 895, L2

\bibitem[Narang et al.(2024)]{narang24}
Narang, M., Manoj, P., Tyagi, H., Watson, D., Megeath, T., et al. 2024, \href{https://doi.org/10.3847/2041-8213/ad1de3} \apjl, 962, L16

\bibitem[Navarro et al.(2025)]{navarro25}
Navarro, M. G., Nisini, B., Giannini, T., Kavanagh, P., et al. 2025, \href{https://doi.org/10.3847/1538-4357/ae1f8f} \apj, 995, 199

\bibitem[Neufeld et al.(2007)]{neufeld07}
Neufeld, D. A., Hollenbach, D. J., Kaufman, M. J., et al. 2007, \href{https://doi.org/10.1086/518857} \apj, 664, 890

\bibitem[Neufeld et al.(2009)]{neufeld09}
Neufeld, D. A., Nisini, B., Giannini, T., et al. 2009, \href{https://doi.org/10.1088/0004-637X/706/1/170} \apj, 706, 170

\bibitem[Nisini et al.(2005)]{nisini05}
Nisini, B., Bacciotti,F., Giannini, T., Massi, F., et al. 2005, \href{https://doi.org/10.1051/0004-6361:20053097} \aap, 441, 159

\bibitem[Nisini et al.(2016)]{nisini16}
Nisini, B., Giannini, T., Antoniucci, S., Alcala, J. M., et al. 2016, \href{http://dx.doi.org/10.1051/0004-6361/201628853} \aap, 595, A76

\bibitem[Nisini et al.(2024)]{nisini24}
Nisini, B., Navarro, M. G., Giannini, T., Antoniucci, S., et al. 2024, \href{https://doi.org/10.3847/1538-4357/ad3d5a} \aap, 967, 168

\bibitem[\"{O}berg et al.(2011)]{oberg11}
\"{O}berg, K. I., Boogert, A. C., Pontoppidan, K. M., van den Broek, S., et al. 2011, \href{http://doi.org/10.1088/0004-637X/740/2/109} \apj, 740, 109

\bibitem[Offner \& Arce (2014)]{offner14}
Offner, S., Arce, H., 2014, \href{http://dx.doi.org/10.1088/0004-637X/784/1/61} \apj, 784, 61

\bibitem[Ohashi et al.(1997)]{ohashi97}
Ohashi, N., Hayashi, M., Ho, P., \& Momose, M., 1997, \href{https://doi.org/10.1086/303533} \apj, 475, 211

\bibitem[Ohashi et al.(2023)]{ohashi23}
Ohashi, N., Tobin, J., Jorgensen 2023, \href{https://doi.org/10.3847/1538-4357/acd384} \apj, 951, 8

\bibitem[Oya et al.(2015)]{oya15}
Oya, Y., Sakai, N., Leefloch, B., Lopez-Sepulcre, A., et al. 2015, \href{http://doi.org/10.1088/0004-637X/812/1/59} \apj, 812, 59

\bibitem[Pascucci et al.(2009)]{pascucci09}
Pascucci, I., \& Sterzik, M. 2009, \href{https://doi.org/10.1088/0004-637X/702/1/724} \apj, 702, 724

\bibitem[Pascucci et al.(2020)]{pascucci20} 
Pascucci, I., Banzatti, A., Gorti, U., Fang, M., et al. 2020, \href{https://doi.org/10.3847/1538-4357/abba3c} \apj, 903, 78

\bibitem[Pascucci et al.(2023)]{pascucci23} 
Pascucci, I., Cabrit, S., Edwards, S., et al. 2023, \href{http://linker.aanda.org/10.1051/0004-6361/202452121/42} PPVII, ASPC Series, Vol.9, 534,

\bibitem[Plunkett et al.(2015)]{plunkett15}
Plunkett, A., Arce, H., Mardones, D., van Dokkum, P., et al. 2015, \href{https://doi.org/10.1038/nature15702} Nature, 527,70

\bibitem[Pontoppidan et al.(2008)]{pont08}
Pontoppidan, K., Boogert, A., Fraser, J., van Dishoeck, E., Blake, G., et al. 2008, \href{https://doi.org/10.1086/533431} \apj, 678, 1005

\bibitem[Pontoppidan et al.(2024)]{pont24}
Pontoppidan, K., Evans, N., Bergner,J., \& Yang, Y-L., 2024, \href{https://doi.org/10.3847/2515-5172/ad303f} Research notes of AAS, 8, 68

\bibitem[Pudritz et al.(2007)]{pudritz07}
Pudritz, R. E., Ouyed, R., Fendt, C., Brandenburg, A., 2007, \href{https://doi.org/10.48550/arXiv.astro-ph/0605597} PPV, Univ. of Arizona press

\bibitem[Pudritz \& Ray (2019)]{pudritz19}
Pudritz, R. E., \& Ray, T. P., 2019, \href{https://doi.org/10.3389/fspas.2019.00054} Frontiers in Astronomy, Vol 6

\bibitem[Raga et al.(2002)]{raga02}
Raga, A. C., Velazquez, P. F., Canto, J., Masciadri, E., 2002, \href{https://doi.org/10.1051/0004-6361:20021180} \aap, 395, 647

\bibitem[Ray et al.(2007)]{ray07}
Ray, T., Dougados, C., Bacciotti, F., Eisl\"{o}ffel, J., \& Chrysostomou, A. 2007, \href{https://ui.adsabs.harvard.edu/link_gateway/2007prpl.conf..231R/doi:10.48550/arXiv.astro-ph/0605597} PPV, Univ. of Arizona press

\bibitem[Ray \& Ferreira(2021)]{ray21}
Ray, T.P. \& Ferreira, J., 2021, \href{https://doi.org/10.1016/j.newar.2021.101615} New Astronomy Reviews, 93, 101615

\bibitem[Ray et al.(2023)]{ray23}
Ray, T.P., McCaughrean, M.J., Caratti o Garatti, A., et al. 2023, \href{https://doi.org/10.1038/s41586-023-06744-8} Nature, 623, E3 

\bibitem[Redaelli et al.(2019)]{redaelli19}
Redaelli, E.,Alves, F. O., Santos, F. P., and Caselli, P., 2019, \href{https://doi.org/10.1051/0004-6361/201936271} \aap, 631, A154 

\bibitem[Reipurth et al.(2004)]{reipurth04}
Reipurth, B., Rodriguez, L., Anglada, G. \& Bally, J., 2004, \href{https://doi.org/10.1086/381062} \aj, 127, 1736

\bibitem[Rieke et al.(2015)]{rieke15}
Rieke, G., Wright, G., Boker, T., Bouwman, J., et al. 2015, \href{https://doi.org/10.1086/682252}\pasp, 127, 584

\bibitem[Rocha et al.(2024)]{rocha24}
Rocha, W. R. M., van Dishoeck, E., Ressler, M. R., van Gelder, M. L., Slavicinska, K., Brunken, N. G. C., et al. 2024, \href{https://doi.org/10.1051/0004-6361/202348427} \aap, 683, A124

\bibitem[Segura-Cox et al.(2015)]{cox15}
Segura-Cox, D., Looney, L., Stephens, I., Fernandez-Lopez, M., Kwon, W., et al. 2015, \href{http://doi.org/10.1088/2041-8205/798/1/L2} \apjl, 798, L2

\bibitem[Shang et al.(2002)]{shang02}
Shang H., Glassgold, A. E., Frank, Shu, Susana, L., 2002, \href{https://doi.org/10.1086/324197} \apj, 564, 853

\bibitem[Shang et al.(2006)]{shang06}
Shang, H., Allen, A., Li, Z.-Y., et al. 2006, \href{https://doi.org/10.1086/506513} \apj, 649, 845

\bibitem[Shang et al.(2007)]{shang07}
Shang H., Li Z.-Y., Hirano N., 2007, \href{https://ui.adsabs.harvard.edu/abs/2007prpl.conf..261S/abstract} PPV, Univ. of Arizona Press, 261

\bibitem[Shang et al.(2010)]{shang10}
Shang H., Glassgold A. E., Wei-Chieh, L., et al. 2010, \href{http://doi.org/10.1088/0004-637X/714/2/1733} \apj, 714, 1733

\bibitem[Sheehan et al.(2022)]{sheehan22}
Sheehan, P. D., Tobin, J. J., Li, Z.-Y., van't Hoff, M., et al. 2022, \href{https://doi.org/10.3847/1538-4357/ac7a3b} \apj, 934, 95

\bibitem[Shu et al.(1994)]{shu94}
Shu F., Najita J., Ostriker E., Wilkin F., Ruden S., Lizano S., 1994, \href{https://doi.org/10.1086/174363} \apj, 429, 781

\bibitem[Shu et al.(2007)]{shu07}
Shu F. H., Galli D., Lizano S., Glassgold A. E., Diamond P. H., 2007, \href{https://doi.org/10.1086/519678} \apj, 665, 535

\bibitem[Slavicinska et al.(2025)]{slavic25}
Slavicinska, K., Tychoniec, L., Gabriela Navarro, G., et al. 2025, \href{https://doi.org/10.3847/2041-8213/addb45}\apjl, 986, L19

\bibitem[Sternberg \& Dalgarno (1989)]{sternberg89}
Sternberg, A., \& Dalgarno, A., 2007, \href{https://ui.adsabs.harvard.edu/link_gateway/1989ApJ...338..197S/doi:10.1086/167193} \apj, 338, 197

\bibitem[Takami et al.(2006)]{takami06}
Takami, M., Chrysostomou, A., Ray, T. P., Davis, C. J., et al. 2006, \href{https://doi.org/10.1086/500352} \apj, 641, 357

\bibitem[Tabone et al.(2017)]{tabone17}
Tabone, B., Cabrit, S., Bianchi, E., Ferreira, J., et al. 2017, \href{https://doi.org/10.1051/0004-6361/201731691} \aap, 607, L6

\bibitem[Tamura et al.(1996)]{tamura96}
Tamura, M., Ohashi, N., Hirano, N., Itoh, Y., et al. 1996, \href{https://doi.org/10.1086/118164} \aj, 112, 5

\bibitem[Tazaki et al.(2025)]{tazaki25}
Tazaki, R., Menard, F., Duchene, G., Villenave, M., et al. 2025, \href{https://doi.org/10.3847/1538-4357/ad9c6f} \apj, 980, 49

\bibitem[Tobin et al.(2008)]{tobin08}
Tobin, J. J., Hartmann, L., Calvet, N., D'Alessio, P., 2008, \href{https://doi.org/10.1086/587683} \apj, 679, 1364

\bibitem[Tobin et al.(2010)]{tobin10}
Tobin, J. J., Hartmann, L., Loinard, L., 2010, \href{http://doi.org/10.1088/2041-8205/722/1/L12} \apj, 722, L12

\bibitem[Tobin et al.(2011)]{tobin11}
Tobin, J. J., Hartmann, L., Chiang, H-F., Looney, L., 2011, \href{http://doi.org/10.1088/0004-637X/740/1/45} \apj, 740, 45

\bibitem[Tobin et al.(2012)]{tobin12}
Tobin, J. J., Hartmann, L., Chiang, H-F., Wilner, D., 2012, \href{http://doi.org/10.1038/nature11610} Nature, 679, 1364

\bibitem[Tobin et al.(2013)]{tobin13}
Tobin, J. J., Hartmann, L., Chiang, H-F., Wilner, D., 2013, \href{http://doi.org/10.1088/0004-637X/771/1/48} \apj, 771, 48

\bibitem[Tomisaka (1998)]{tomisaka98}
Tomisaka, K., 1998, \href{http://doi.org/10.1086/311504} \apj, 502, L163

\bibitem[Towner et al.(2023)]{towner23}
Towner, A. M., Ginsburg, A., Dell’Ova, P., Gusdorf, A., et al. 2023, \href{https://doi.org/10.3847/1538-4357/ad0786} \apj, 960, 48

\bibitem[Tychoniec et al.(2024)]{tycho24}
Tychoniec, L., van Gelder, M. L., van Dishoeck, E., Francis, L., Rocha, W., et al. 2024, \href{https://doi.org/10.1051/0004-6361/202348889} \aap, 687, A36

\bibitem[van Dishoeck et al.(2025)]{dishoeck25}
van Dishoeck, E., Tychoniec, L., Rocha, W., Slavicinska, K., Francis, L., et al. 2025, \href{https://doi.org/10.1051/0004-6361/202554444} \aap, 699, A361

\bibitem[van Gelder et al.(2024)]{gelder24}
van Gelder, M. L., Francis, L., van Dishoeck, E., Tychoniec, L., Ray, T. P., et al. 2024, \href{https://doi.org/10.1051/0004-6361/202451967} \aap, 692, A197

\bibitem[van't Hoff et al.(2023)]{vanthoff23}
van't Hoff, M. L. R., Tobin, J. J., Li, Z.-Y., et al. 2023, \href{https://doi.org/10.3847/1538-4357/accf87} \apj, 951, 10

\bibitem[Wright et al.(2023)]{wright23}
Wright, G., Rieke, G., Glasse, A., Ressler, M., Garcia Marin, M., et al. 2023, \href{https://doi.org/10.1088/1538-3873/acbe66} \pasp, 135, 048003

\bibitem[Wu et al.(2004)]{wu04}
Wu Y., Wei Y., Zhao M., Shi Y., et al. 2004, \href{https://doi.org/10.1051/0004-6361:20035767} \aap, 426, 503

\bibitem[Yang et al.(2022)]{yang22}
Yang, Y.-L., Green, J. D., Pontoppidan, K. M., et al. 2022, \href{https://doi.org/10.3847/2041-8213/aca289} \apj, 941, L13

\bibitem[Yildiz et al.(2015)]{yildiz15}
Yildiz, U. A., Kristensen, L. E., van Dishoeck, E. F., et al. 2015, \href{http://doi.org/10.1051/0004-6361/201424538} \aap, 576, A109

\bibitem[Zhang et al.(2016)]{zhang16}
Zhang, Y., Arce, H. G., Mardones, D., et al. 2016, \href{http://doi.org/10.3847/0004-637X/832/2/158} \apj, 832, 158

\bibitem[Zhou et al.(1996)]{zhou96}
Zhou, S., Evans, N. \& Wang, Y., 1996, \href{https://doi.org/10.1086/177510} \apj, 466, 296
%
\end{thebibliography}
\end{document}